\documentclass[12pt,a4paper]{article}
\pdfoutput=1

\usepackage{amsmath}
\usepackage{color}
\usepackage{amsfonts}
\usepackage{amssymb}
\usepackage{graphicx}
\usepackage{geometry}
\usepackage{amssymb,epsfig,subfigure}
\usepackage{hyperref}
\usepackage{comment}
\usepackage[font=footnotesize]{caption}

\makeatletter
\renewcommand\section{\@startsection {section}{1}{\z@}%
                                 {-3.5ex \@plus -1ex \@minus -.2ex}
                                   {2.3ex \@plus.2ex}%
                                   {\normalfont\large\bfseries}}
\renewcommand\subsection{\@startsection{subsection}{2}{\z@}%
                                   {-3.25ex\@plus -1ex \@minus -.2ex}%
                                     {1.5ex \@plus .2ex}%
                                     {\normalfont\bfseries}}
\renewcommand\subsubsection{\@startsection{subsubsection}{3}{\z@}%
                                   {-3.25ex\@plus -1ex \@minus -.2ex}%
                                     {1.5ex \@plus .2ex}%
                                     {\normalfont\itshape}}
\makeatother

\def\pplogo{\vbox{\kern-\headheight\kern -29pt
\halign{##&##\hfil\cr&{\ppnumber}\cr\rule{0pt}{2.5ex}&\ppdate\cr}}}
\makeatletter
\def\ps@firstpage{\ps@empty \def\@oddhead{\hss\pplogo}%
  \let\@evenhead\@oddhead 
}
\def\maketitle{\par
 \begingroup
 \def\thefootnote{\fnsymbol{footnote}}
 \def\@makefnmark{\hbox{$^{\@thefnmark}$\hss}}
 \if@twocolumn
 \twocolumn[\@maketitle]
 \else \newpage
 \global\@topnum\z@ \@maketitle \fi\thispagestyle{firstpage}\@thanks
 \endgroup
 \setcounter{footnote}{0}
 \let\maketitle\relax
 \let\@maketitle\relax
 \gdef\@thanks{}\gdef\@author{}\gdef\@title{}\let\thanks\relax}
\makeatother

\numberwithin{equation}{section}

\newcommand\eea{\end{eqnarray}}
\newcommand\bea{\begin{eqnarray}}

\def\beq{\begin{equation}}
\def\eeq{\end{equation}}

\newcommand{\be}{\begin{equation}}
\newcommand{\ee}{\end{equation}}
\newcommand{\ba}{\begin{align}}
\newcommand{\ea}{\end{align}}
\newcommand{\bg}{\begin{gather}}
\newcommand{\eg}{\end{gather}}
\newcommand{\bseq}{\begin{subequations}}
\newcommand{\eseq}{\end{subequations}}

\begin{document}
\setcounter{page}0
\def\ppnumber{\vbox{\baselineskip14pt
}}
\def\ppdate{
} \date{}

\author{Horacio Casini, Eduardo Test\'e, Gonzalo Torroba\\
[7mm] \\
{\normalsize \it Centro At\'omico Bariloche and CONICET}\\
{\normalsize \it S.C. de Bariloche, R\'io Negro, R8402AGP, Argentina}
}

\bigskip
\title{\bf  Modular Hamiltonians on the null plane \\ 
\vskip 2mm
and the Markov property of the vacuum state
\vskip 0.5cm}
\maketitle

\begin{abstract}
We compute the modular Hamiltonians of regions having the future horizon lying on a null plane. For a CFT this is equivalent to regions with boundary of arbitrary shape lying on the null cone. These Hamiltonians have a local expression on the horizon formed by integrals of the stress tensor. We prove this result in two different ways, and show that the modular Hamiltonians of these regions form an infinite dimensional Lie algebra. The corresponding group of unitary transformations moves the fields on the null surface locally along the null generators with arbitrary null line dependent velocities, but act non locally outside the null plane. We regain this result in greater generality using more abstract tools on algebraic quantum field theory. Finally, we show that modular Hamiltonians on the null surface satisfy a Markov property that leads to the saturation of the strong sub-additive inequality for the entropies and to  the strong super-additivity of the relative entropy.
\end{abstract}
\bigskip

\newpage

\tableofcontents

\vskip 1cm

\section{Introduction}\label{sec:intro}

In recent years there has been much interest in the statistical properties of the vacuum state. If we restrict attention to a spatial region $V$, the vacuum state is represented by a density matrix $\rho_V$. Its entropy $S(V)$ --which is produced  by entanglement with the complementary region $\bar{V}$-- has important applications in quantum field theory (QFT), holography and black hole physics. A more natural way to describe the density matrix is through its modular Hamiltonian $H_V=-\log \rho_V$. In terms of the modular Hamiltonian the density matrix has a ``thermal" like form $\rho_V=e^{-H_V}$. In the same spirit we can define a one-parameter group of unitaries $U_V(s)=e^{-i s H_V }=\rho^{i s}$ that are internal ``time"  transformations on the algebra of operators in $V$. This is called the modular group or the modular flow corresponding to $V$. Because of causality, the algebra of operators on $V$ also contains all localized operators in the causal development of the spatial region $V$. In this paper we often do not make distinctions between a spatial region and its causal development. 

Modular Hamiltonians are very interesting objects that encode all properties of the vacuum in a region. In particular,  for a state $\rho^1$ that is a small deviation of $\rho^0$ the modular Hamiltonian $H=-\log \rho^0$ gives a way to compute the variation of the entropy
\be
\Delta S=S(\rho^1)-S(\rho^0)=\langle H\rangle_1-\langle H\rangle_0=\Delta \langle H\rangle\,.
\ee
This is called the first law of entanglement entropy in analogy with the first law of thermodynamics. This property has been the subject of much of the recent interest and applications of modular Hamiltonians; see for example \cite{Blanco:2013joa,Wong:2013gua, Herzog:2014fra, Jafferis:2014lza, Lashkari:2015dia, Jafferis:2015del, Faulkner:2015csl, Cardy:2016fqc, Sarosi:2016oks}. In a similar way, the modular Hamiltonian enters in the calculation of the relative entropy 
\be\label{eq:Srel0}
S(\rho^1||\rho^0)=\textrm{tr}(\rho^1\log \rho^1-\rho^1\log\rho^0)
\ee 
between two arbitrary states. We can write
\be\label{eq:Srel1}
S(\rho^1||\rho^0)=\Delta \langle H\rangle-\Delta S\,. 
\ee
This quantity has an operational interpretation as a statistical distance between the states, and is a central quantity in quantum information theory. In particular it is positive and decreasing under taking subsystems. Among applications of the relative entropy to QFT we can mention the proof of the Bekenstein bound \cite{Casini:2008cr, Blanco:2013lea}, the quantum averaged null energy condition (ANEC) \cite{Faulkner:2016mzt, Hartman:2016lgu}, and results on the monotonicity of renormalization group (RG) flows \cite{Casini:2016fgb, Casini:2016udt}.

In the mathematical literature modular Hamiltonians have been studied since the 70's. They play a structural role in the algebraic formulation of QFT \cite{Haag:1992hx,Borchers.unicidad}. A particularly important result, intimately related to the CPT theorem, is the fact that the modular Hamiltonian for the Rindler wedge (the region $x^1\ge |x^0|$ in Minkowski space) is given by 
\be
H=2\pi K_1\,, \label{wwr}
\ee 
where $K_1$ is the boost generator in the direction of $x^1$ \cite{Bisognano:1975ih}. Hence, the modular Hamiltonian for the wedges has an expression as an integral of the stress tensor for any QFT. The same holds for spheres in conformal field theories (CFT) because spheres are conformally related to planes \cite{Hislop:1981uh, Casini:2011kv}. This expression in terms of local operators is rather surprising from the point of view of quantum information theory (QIT). One would naively expect the logarithm of the density matrix to be a completely non local operator in the region.\footnote{See \cite{Arias:2016nip} for a discussion on local terms in $H$ for general regions in QFT.}    
 The particularity of these cases is that there is a space-time symmetry with time-like killing vector keeping the region fixed. 

In this paper we study modular Hamiltonians and modular flows for regions with arbitrary shape lying on the null cone in a CFT. Equivalently, these regions can be conformally transformed and thought as regions having as future horizon a sector of a null plane. We then extend our results to relevant deformations of CFTs.

\begin{figure}[h!]
\begin{center}  
\includegraphics[width=0.6\textwidth]{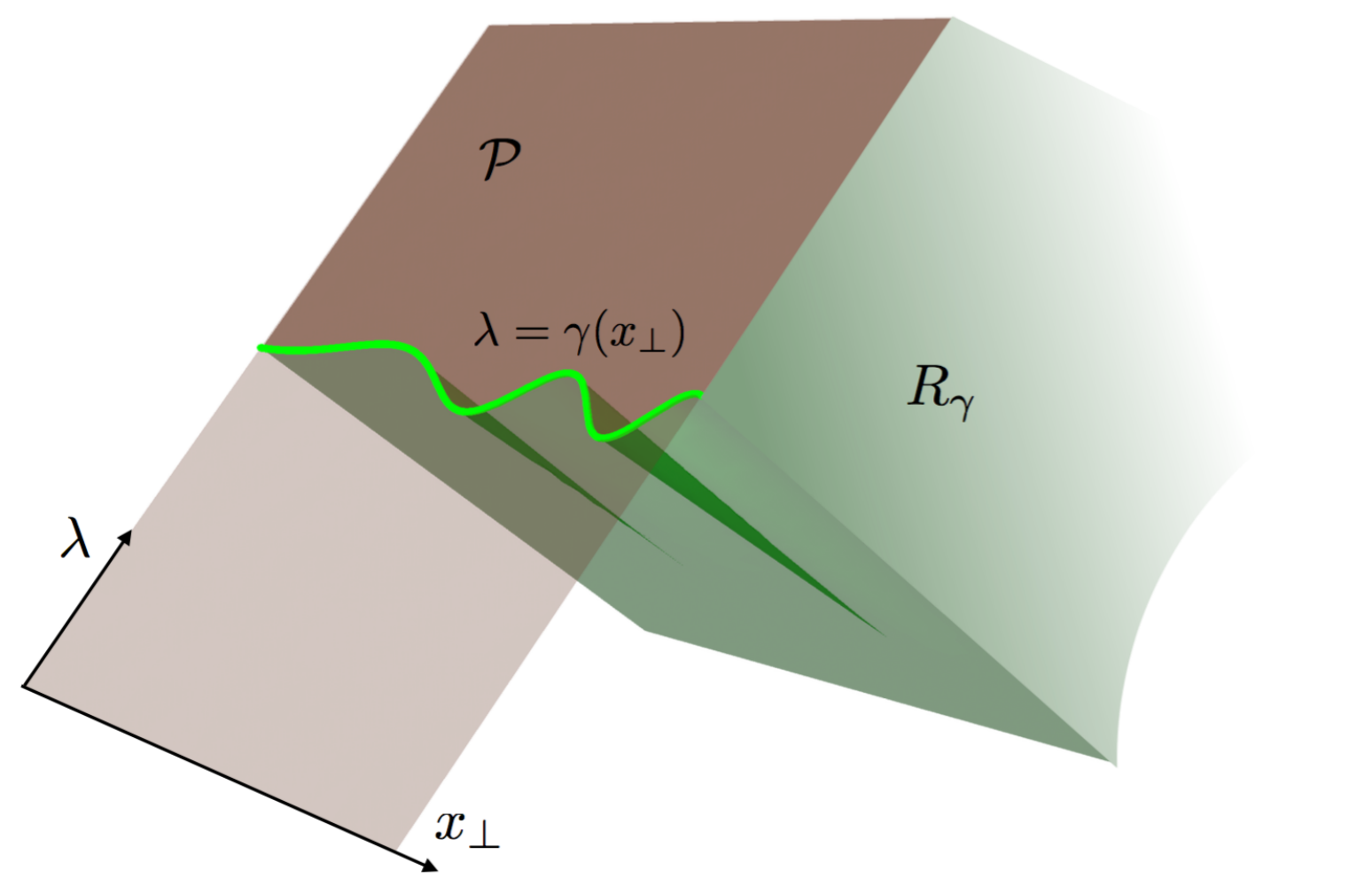}
\captionsetup{width=0.9\textwidth}
\caption{Setup of the work: null plane $\mathcal P$ parallel to $\xi = (1,1,0,\cdots)$, with an arbitrary curve $\gamma(\lambda)$. We will determine the modular Hamiltonians associated to regions $R_\gamma$.}
\label{region}
\end{center}  
\end{figure}  

The setup is illustrated in Fig. \ref{region}. We consider the null plane ${\cal P}$ through the origin in Minkowski space, parallel to the null vector $\xi=(1,1,0...)$. We take the coordinates on the null plane as $(\lambda,x^\perp)$ with $x=\lambda \xi+x^\perp$, $x^\perp=(0,0,x^2,...x^{d-1})$. Let us consider a $d-2$ dimensional spatial surface $\gamma$ on this null plane, given by the equation $\lambda=\gamma(x^\perp)$. We are going to take $\gamma$ to be continuous, extending all the way to infinity in the coordinates $x^\perp$, and dividing the null plane in two.
We are interested in causal regions $R_\gamma$ of the space-time having as future horizon all the points on ${\cal P}$ at the future of $\gamma$. That is, all points that are spatial to all points spatial to the future of $\gamma$ in ${\cal P}$.    Simplifying the notation, we will sometimes label these space-time regions, and the corresponding algebra of operators generated by fields smeared in the region, with the same name $\gamma$. We denote by $\bar{\gamma}$ the complementary region (all points spatially separated from $\gamma$) that has a past horizon containing the points on ${\cal P}$ to the past of $\gamma$. This is associated to the algebra of operators that commute with the ones in $\gamma$.

We prove that the modular Hamiltonian of $\gamma$ on the future horizon is given by
\be
H_\gamma=2\pi \int d^{d-2} x^\perp\, \int_{\gamma(x^\perp)}^\infty d\lambda\, (\lambda-\gamma(x^\perp)) T_{\lambda\lambda}(\lambda,x^\perp)\,. \label{hmodi}
\ee    
 This generalizes (\ref{wwr}) to any region having spatial boundary on the null plane. Here again we encounter a local expression in terms of the stress tensor; it is in fact of the form of the Rindler result on each null generator of the surface individually. For the case of wedges, when $\gamma$ is a plane, this corresponds to the flux over the null surface of a conserved current, and we can write the modular Hamiltonian, proportional to a boost operator, in any other spatial surface as an integral of the stress tensor. In general this is not the case for other $\gamma$'s --we would have a non local expression on Cauchy surfaces other than the null surface. 
   
The result (\ref{hmodi}) was proved for free fields in \cite{Wall:2011hj} (see also \cite{Bousso:2014uxa}), and for small deformations away from planes in \cite{Faulkner:2016mzt}. Furthermore, in the context of the conjectured quantum null energy condition (QNEC) \cite{Bousso:2015mna, Bousso:2015wca}, this formula was obtained by \cite{Koeller:2017njr} for the class of theories that saturate the QNEC. From this point of view, our results imply that the QNEC is saturated for a small deformation of the vacuum state in QFT.

It is often convenient to define the full modular Hamiltonian for a region $V$ as 
\be
\hat{H}_V=H_V-H_{\bar{V}}\,.
\ee
In contrast to $H_{V}$ this has support on all space and always annihilates the vacuum, $\hat{H}_V |0\rangle=0$. We have simply
\be
\hat{H}_\gamma=2\pi \int d^{d-2} x^\perp\, \int_{-\infty}^\infty d\lambda\, (\lambda-\gamma(x^\perp)) T_{\lambda\lambda}(\lambda,x^\perp)\,. \label{hmodi1}
\ee 
We will show the modular flows generated by these operators $U(s)=e^{-i s \hat{H}_\gamma}$ move all other regions on the null plane in a precise geometrical way. 

We prove these results in different ways. After reviewing conformal transformations and some known results for modular Hamiltonians in a CFT in Section \ref{sec:conf}, we give a first proof in Section \ref{sec:ope}. This is a direct calculation that uses the replica trick and the operator product expansion (OPE) for twisting operators separated along a null direction. In this proof we follow closely the work \cite{Bousso:2014uxa}. 

Based on some useful mathematical properties reviewed in Section \ref{sec:mate}, in Section \ref{sec:algebra} we give a second proof that starts by computing the algebra of the operators (\ref{hmodi1}). This forms an infinite-dimensional Lie algebra with commutators 
\be
[\hat{H}_{\gamma_1},\hat{H}_{\gamma_2}]=2 \pi i (\hat{H}_{\gamma_1}-\hat{H}_{\gamma_2})\,.\label{alge}
\ee
 In contrast to ordinary space-time symmetries, it moves operators locally on the null surface but not outside it.  Using results on algebraic QFT and the averaged null energy condition (ANEC) proved in \cite{Faulkner:2016mzt, Hartman:2016lgu}, we show that the operators (\ref{hmodi1}) are in fact the full modular Hamiltonians for the regions $R_\gamma$. 

In Section \ref{sec:algebraic} we derive the algebra of modular Hamiltonians and the local action of modular flows from a very general algebraic perspective that does not use any of the standard techniques of CFT. In particular we do not need to use the stress tensor in an explicit way. However, because of this same generality this approach does not give an explicit formula for modular Hamiltonians. Our results are heavily based on the investigations about half-sided modular inclusions that where developed in the algebraic approach to QFT \cite{wiesbrock1993half,Borchers.unicidad}.   

The form of the modular Hamiltonians (\ref{hmodi}) leads to the following equation for two intersecting regions $\gamma_1$, $\gamma_2$, 
\be\label{markovsinhat} 
H_{\gamma_1}+H_{\gamma_2}-H_{\gamma_1\cap \gamma_2}-H_{\gamma_1\cup \gamma_2}=0\,.
\ee
This equation follows trivially on each null line separately. In fact, any of our thee approaches leads to this equation, and in particular we can prove it in a more general context based on the theory of half-sided modular inclusions for algebras. In Section \ref{sec:markov} we show that this is equivalent to a Markov property for the vacuum state on these regions. This has the interesting consequence that the vacuum entropies saturate the strong sub-additive entropy  inequality,
\be
S_{\gamma_1}+S_{\gamma_2}-S_{\gamma_1\cap \gamma_2}-S_{\gamma_1\cup \gamma_2}=0\,.
\ee
 This also leads to a strong super-additive property of the relative entropy between any state and the vacuum state for these regions satisfying the Markov property. These elements are central to a new proof of the a-theorem about irreversibility of the renormalization group in $d=4$ that is presented in \cite{Casini:2017vbe}. 

\section{The plane, the cone, and the cylinder}
\label{sec:conf}

\begin{figure}[h!]
\begin{center}  
\includegraphics[scale=0.65]{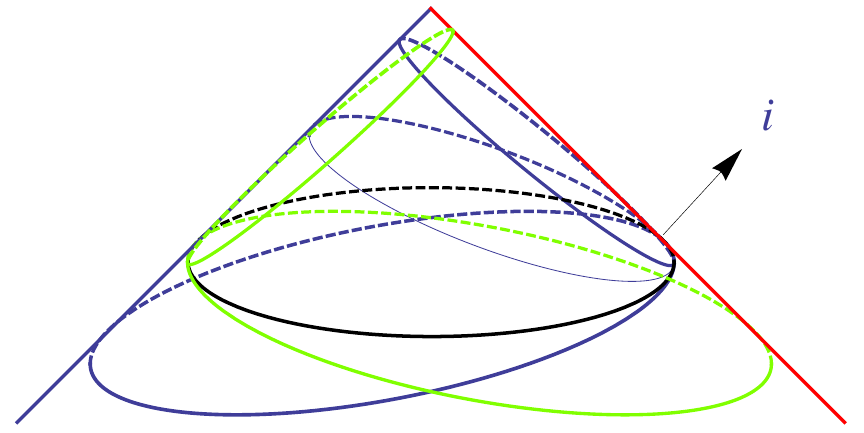}
\captionsetup{width=0.9\textwidth}
\caption{The past null cone in Minkowski space. The spheres in the cone passing though the point $i$ (blue) are mapped to planes on the null plane with the conformal transformation (\ref{maa}). All other spheres (green) are mapped to parabolic regions on the null plane. The point $i$ and the red null line are mapped to infinity.}
\label{cono}
\end{center}  
\end{figure}

For a CFT in Minkowski space a natural setting of the problem is for regions with arbitrary boundaries on the past light cone of a point.  However, for simplicity, we can use regions on the null plane since this null cone can be mapped to the null plane by a conformal transformation. We will work in the metric signature $(+--...-)$. The conformal mapping from Minkowski space with coordinates $X^\mu$ to Minkowski with coordinates $x^\mu$
\be
x^\mu=2\frac{X^\mu-(X\cdot X)C^\mu}{1-2(X\cdot C)+(X\cdot X)(C\cdot C)}-R^2 C^\mu\,,\ \ \ \ C^\mu\equiv(0,1/R,\vec{0})\ ,\label{maa}
\ee 
maps the Rindler wedge $X^\pm=X^1\pm X^0\ge 0$ into the causal diamond of a sphere centered at the origin, $x^\pm=r\pm x^0\le R$. In particular, the origin $X^\mu=0$ is mapped into the point on the left of the sphere $(0,-R,\vec{0})$, the future and past horizons of Rindler are mapped to the ones of the sphere. The point on the right of the sphere $i=(0,R,\vec{0})$ corresponds to spatial infinity in the coordinates $X$ --see Fig. \ref{cono}. All points on the light ray marked on red in the figure also correspond to infinity in the original coordinates.    
Part of the plane $X^-=0$ is mapped to the past null cone of the point $(R,\vec{0})$ which is seated at the tip of the cone. All spheres passing through $i$ correspond to planes on the null surface $X^-=0$. Other spheres on the null cone correspond to parabolas 
\be
\lambda=\lambda_0+a (x^\perp-x^\perp_0)^2  \label{parabola}
\ee 
in the original null plane with symmetry axes pointing in the $(1,1,\vec{0})$ direction. These parabolas can have arbitrarily large $|a|$, making them as much aligned to a null line as we want. 

Let us review in the language of the null plane what is known about modular Hamiltonians and modular flows of a CFT.  
The modular Hamiltonian of planes $\gamma$ are well known to have the form (\ref{hmodi}) when written on the null plane. They are the flux of the conserved current corresponding to a boost generator. 
The modular Hamiltonians of spheres are obtained by conformal mappings of modular Hamiltonians of wedges \cite{Hislop:1981uh, Casini:2011kv}. When a sphere is transformed to a parabola in the plane, we find the same expression (\ref{hmodi}) where now $\gamma$ is a parabola. 
This can be computed directly, but we can more simply argue as follows. We obtain the parabola by starting with a half null plane, a mapping to the cone, a boost and a rotation that keeps the cone invariant, and a transformation back to the null plane. Hence we have a conformal transformation of the half null plane to the parabola. This maps any semi-infinite null line to another semi-infinite null line, and then it must be a translation and a dilatation on the null line. However, a dilatation will change the velocity of the modular flow, in particular near the end point of the semi-infinite null line, where the modular flow must be Rindler-like. Then we conclude the transformation is just a translation for each independent null line. This gives the result (\ref{hmodi}) for parabolas as well. The corresponding modular flows act locally and in particular they transform each null ray in itself as can be deduced from the form of (\ref{hmodi}). Therefore, the modular flows of parabolas or planes $\gamma_1$ act on any other surface $\gamma_2$ as
\be
U_{\gamma_1}(-s) \gamma_2 U_{\gamma_1}(s) =e^{2\pi s}(\gamma_2-\gamma_1)+\gamma_1\,.\label{standard00}
\ee         

\begin{figure}[t]
\begin{center}  
\includegraphics[scale=0.33]{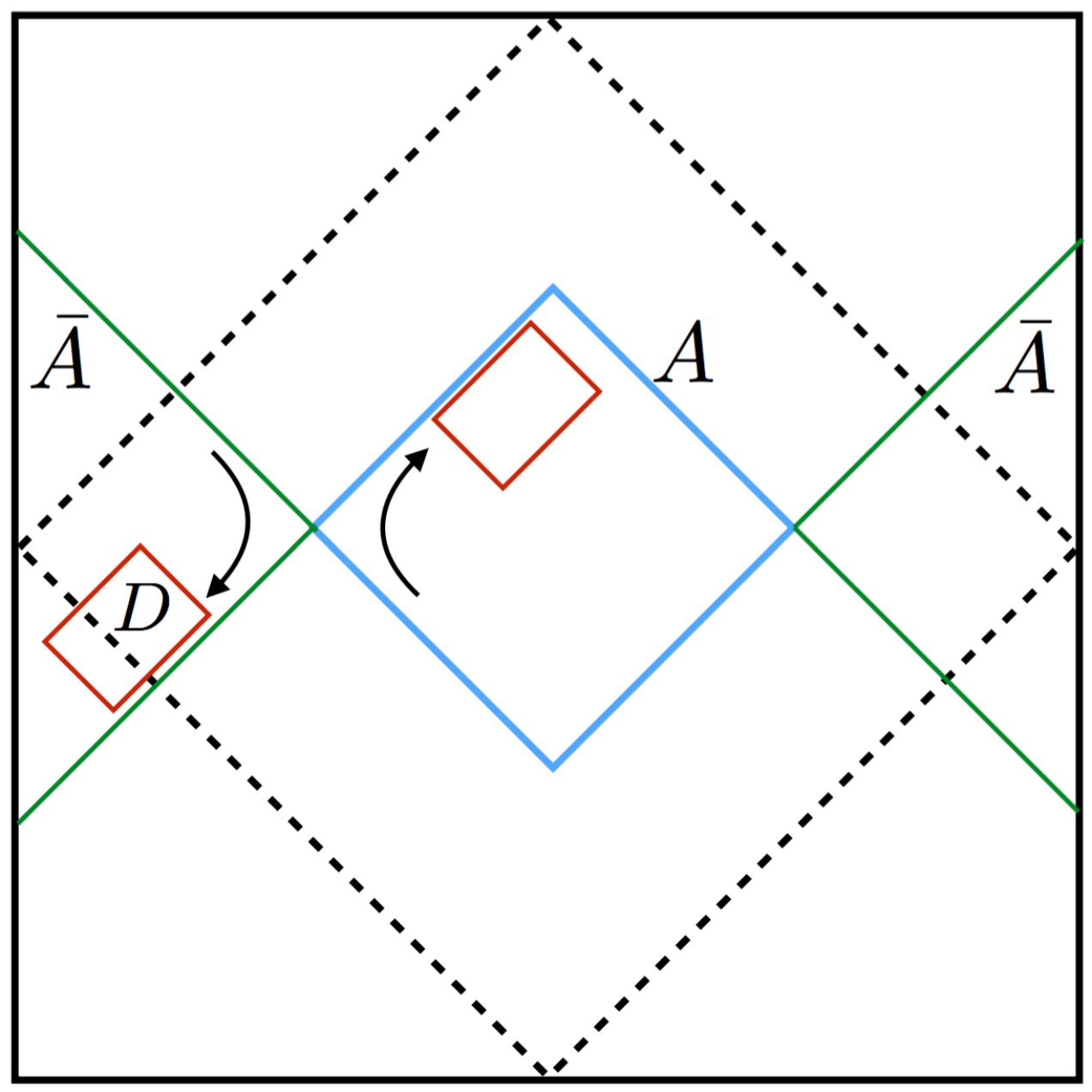}
\captionsetup{width=0.85\textwidth}
\caption{The cylinder where the two vertical edges are identified. Minkowski space is conformal to the diamond shown with dashed lines. The modular flow of a double cone $A$ inside Minkowski space moves regions inside $A$ towards the future tip of $A$ (for positive modular parameter) and regions inside the complement ($D$ in the figure) towards the past tip of the complement in the cylinder. For some finite modular parameter a point in the complement will cross the past null boundary of Minkowski space making the flow in Minkowski non local. However, the flow is still local in the cylinder.  }
\label{cilindro}
\end{center}  
\end{figure}

There is however a subtle point in this geometry. In the mapping (\ref{maa}) the points on the null plane $X^-=0$
 are mapped to infinity for finite negative values of $\lambda=X^0=X^1$ of this null plane, 
\be
\lambda=-\frac{(x^\perp)^2}{2 R}-\frac{R}{2}\,.\label{pariy}
\ee 
Hence, for example, some of the parabolas in (\ref{parabola}) that cross the surface (\ref{pariy}) do not come from spheres in the null cone. Another disturbing feature (illustrated in Fig. \ref{cilindro}) is that the modular flow of a sphere acting on a point $p$ of the null cone below the sphere will push it down the cone but at some finite value of the modular parameter it will reach infinity. Hence, in Minkowski space where the cone lives, the modular flow of the sphere does not act locally on $p$ after this value of the modular parameter (though the modular flow continues to exist as a unitary transformation for all values of the parameter). 

These issues are closely related to the fact that the conformal group $SO(2,d)$ does not act on Minkowski space since special conformal transformations can map points to infinity. Instead, the CFT is naturally defined on a cylinder, obtained by compactifying Minkowski space \cite{Witten:1998qj, Weinberg:2010fx}. In Fig. \ref{cilindro}, Minkowski space is shown as the diamond (drawn with the dashed lines) inside the cylinder. In the cylinder there is no difference any more between wedges and sphere diamonds. The causal complement of a spherical diamond is another spherical diamond. The modular flow of the diamond $A$ pushes regions inside $A$ towards the tip of the future horizon, while it pushes regions on the complement of $A$ towards the tip of the past horizon of the complement.
  In doing so, at some point the flow pushes points on the complement across the line marking the past null infinity of Minkowski space. A field on a point below this line can still be represented as a combination of operators on Minkowski space since Minkowski space contains a full Cauchy surface for the cylinder. However, it will not be represented as a local operator and the flow will cease to be local. The same unitary flow of operators is represented locally on the cylinder. Therefore we can say that the ``truth" of what happens in Minkowski space once one encounters coordinate singularities in conformal transformations or modular flows can be read off from the cylinder.      

The conformal mapping from Minkowski space with coordinates $t, \vec{x}$  to the cylinder is given by \cite{Candelas:1978gf}
\be\label{eq:Mink-cyl}
ds^2_{Mink}=\omega^{-2}(t_c, \psi) \left(dt_c^2-d\psi^2-\sin^2(\psi) d\Omega^2\right)
\ee
with (define $r=|\vec{x}|$)
\be
t\pm r=\tan\left(\frac{t_c\pm \psi}{2}\right)\,,
\ee
and Weyl factor
\be\label{eq:Weyl}
\omega= \cos t_c+\cos \psi\,.
\ee
This maps the full null plane in Minkowski space to the full past light cone of a point in the cylinder, which coincides with the future light cone of another point. As we move from past infinity to infinity along the null ray in Minkowski space we move from a point $p$ on the spherical cross section of the cylinder into the future until it collides with all other null rays from $p$ at the opposite site on the sphere.   

Hence, we see there is also a technical advantage in working with regions on the null plane rather than the null cone in Minkowski space: we are allowed now to think on all parabolas of the form (\ref{parabola}) without restrictions. 
However, in the null plane version of the regions $\gamma$ we are considering, it would seem there is another disconnected piece of the Cauchy surface lying at future null infinity that is missing in our description of the modular Hamiltonians (\ref{hmodi}). This will be infinitely far apart and the modular Hamiltonian for a theory in Minkowski space should contain another term in that surface. However, for a CFT coming from the cylinder we see this term does not exist, since the Cauchy surface is complete there. We will continue to use the null plane description for simplicity.

\section{Modular Hamiltonians on the null plane I. OPE expansion}
\label{sec:ope}

In this section we present our first proof of (\ref{hmodi}). This is based on the replica trick and the OPE of twist operators in the null limit, studied in \cite{Bousso:2014uxa}.\footnote{A similar analysis of OPE expansions of twist operators was done with other purposes in \cite{Cardy.esferaslejanas}.} Most of our analysis will be for CFTs, and hence valid on the plane or the cone; towards the end of the section we extend the result (\ref{hmodi}) to relevant deformations of CFTs. This applies to the null plane, but the map to the cone is no longer possible since it uses conformal transformations.

\subsection{Operator product expansion for twist fields}

Let us first review the results of \cite{Bousso:2014uxa}. We will also resolve a remaining issue in that work regarding singular contributions from fields with twist $\tau<d-2$.

Writing $x^\pm=x^0\pm x^1$,\footnote{That is, the relation with the previously defined parameter $\lambda$ is $x^+=2 \lambda$.} we are interested in the modular Hamiltonian for a strip of width $\Delta x^+$ on the null plane $\mathcal P$ above; we will later on take the limit $\Delta x^+ \to \infty$. This can be calculated using the replica trick as follows. The $n$-th R\'enyi entropy for a spatial strip is determined by an expectation value of two twist operators (one on each boundary of the strip) in an euclidean theory given by $n$ copies of the CFT \cite{Calabrese:2009qy}. In the limit of small width, this correlator admits an operator product expansion (OPE) in terms of local operators of the replicated CFT. The euclidean OPE is then continued to Minkowski space; taking the null limit $\Delta x^- \to 0$, $\Delta x^+$ fixed, gives \cite{Bousso:2014uxa}
\bea\label{eq:OPE1}
D_n(\Delta x) D_n(0) \sim \exp \left[(1-n) \sum_{i=1}^n\sum_k \int d^{d-2} x^\perp \int_0^{\Delta x^+} dx^+ \,\frac{1}{|\Delta x^- \,x^+|^{\frac{1}{2}(d-2-\Delta_k+s_k)}} (x^+)^{s_k-1}
 \right. \nonumber\\
\left.  \times  \,\mathcal O^i_{\Delta_k,s_k}(x^+,x^\perp) +\sum_{i=1}^n\sum_{j=1}^n \cdots \right]\,.\qquad\;\label{twist}
\eea
The integral over $x^+$ reflects the fact that all operators on a null line can contribute to the OPE in the null limit. Translation invariance along the directions parallel to the strip forbid nontrivial dependence on $x^\perp$. The factor $(1-n)$ indicates that this operator goes to the identity for $n\rightarrow 1$.

Here the first sum in the exponent is over the index $i$ of the different copies of the $n$-times replicated space, and we  sum over different operators with index $k$ on each copy.  The missing terms indicated by the ellipsis contain products of fields in different copies.
$\Delta_k$ is the dimension of the operator, and $s_k$ is the spin (boost eigenvalue) on the two-dimensional plane $(x^0, x^1)$. 
 Since twist operators  are spinless and dimensionless, all powers in this expansion are fixed to have total dimension and spin equal to zero on each term. 
We can focus on the leading contribution of maximal spin $\mathcal O_{+ \ldots +}$ since $\Delta x^- \to 0$. The null limit is controlled by the twist $\tau \equiv \Delta -s$. In order to understand better this expression, it is useful to recall that the dilatation and spin eigenvalues (boost $J^{01}$) for $x^+, x^-, x^\perp$ are given respectively by $(\Delta , s) = (-1,-1), (-1,1),(-1,0)$. Hence $\tau(x^+)=0$, $\tau(x^-)=-2$, $\tau(x^\perp)=-1$. Furthermore, the unitarity bound for primary operators of spin $s \ge 1$ (symmetric and traceless) gives $\tau\ge d-2$. Conserved currents have $\Delta=d-1$, $s=1$, and saturate the bound. Operators in representations that are not symmetric and traceless can have $\frac{1}{2}(d-2) < \tau \le d-2$. Finally, scalar operators can have $\frac{1}{2}(d-2) \le \Delta \le d-2$ and hence $\tau$ can also be in this range for $s=0$.\footnote{For a recent discussion on OPEs in the null limit and further references see e.g. \cite{Komargodski:2012ek}.}

The entropy difference from a state $|\psi\rangle$ and the vacuum is given by 
\be
\Delta S=\lim_{n\rightarrow 1} (1-n)^{-1}\log \frac{\langle\psi| D_n(\Delta x) D_n(0)|\psi\rangle}{\langle 0| D_n(\Delta x) D_n(0)|0\rangle}\,. 
\ee
For the calculation of the modular Hamiltonian we need to compute $\Delta \langle H\rangle=\Delta S$ for small deviations of the vacuum state. The knowledge of $\Delta \langle H\rangle$ for any small deviation fixes the modular Hamiltonian  operator  uniquely. 
Then, in the limit $n\rightarrow 1$ 
we have to focus on contributions to the entropy proportional to an operator in the original CFT --these contributions are linear in the density matrix and hence the vacuum-subtracted entropy equals $\Delta \langle H\rangle$. This gives
\be
\Delta \langle H\rangle =\sum_k \int d^{d-2} x^\perp \int_0^{\Delta x^+} dx^+ \,\frac{1}{|\Delta x^- \,x^+|^{\frac{1}{2}(d-2-\Delta_k+s_k)}} (x^+)^{s_k-1}
 \,\mathcal \langle O^i_{\Delta_k,s_k}(x^+,x^\perp)\rangle \,.  \label{HH}
\ee
The terms with ellipsis in (\ref{twist}) will produce nonlinear terms in the density matrix, and while they can contribute to $\Delta S$ for generic states, they do not contribute to the modular Hamiltonian.

We need to constraint the possible operators $\mathcal O$ in the above OPE. Operators with $\tau > d-2$ do not contribute in the null limit $\Delta x^- \to 0$. At $ \tau=d-2$ we generically have conserved spin one currents (which cannot appear because of CPT symmetry) and the stress tensor. For $\frac{1}{2}(d-2) < \tau \le d-2$ we could have, as discussed before, representations that are not symmetric, but these would have to appear in pairs hence giving $\tau > d-2$. The last possibility is then scalar operators (with the same quantum numbers of the vacuum) with $\frac{1}{2}(d-2) \le \Delta \le d-2$. They would lead to a divergent contribution in the OPE and \cite{Bousso:2014uxa} was restricted to theories without such operators. Here we note that we cannot fix the $x^+$ dependence of scalar contributions; equivalently, the integral in (\ref{eq:OPE1}) requires $ s \ge 1$.  They could in principle appear at the boundaries of the strip, but in that case they may be absorbed in the definition of the twist operator itself. The form of the Rindler Hamiltonian eliminates this potential ambiguity in the twist operator. Therefore, scalar operators are also absent in the twist OPE, and only the stress tensor (and its descendants) can appear. In this case,  (\ref{HH}) leads to the modular Hamiltonian \cite{Bousso:2014uxa}
\be\label{eq:strip}
H= 2\pi \int d^{d-2}x^\perp\,\int_0^{\Delta x^+}\,dx^+ \,\Delta x^+\,g(x^+/\Delta x^+)\,T_{++}(x^-=0, x^+, x^\perp)\,.
\ee
The function $g(x)$ comes from summing over descendants and was studied in \cite{Bousso:2014uxa}.

\subsection{Modular Hamiltonian for arbitrary shapes}\label{subsec:shapes}

We now need to take the limit of large $\Delta x^+$ and allow for an arbitrary shape $\gamma(x^\perp)$, as described in Section \ref{sec:intro}. The key result from (\ref{eq:strip}) is the independence of null lines on the plane. Intuitively, this is because the distance between points in two parallel null lines is independent of the position $\lambda$ along the line. This suggests that the factorization between different null rays should continue to hold for arbitrary shapes, allowing us to write the modular Hamiltonian as a sum over the different rays. We will argue, using the twist OPE, that this is indeed correct.

For this, let us go back to (\ref{eq:OPE1}), now allowing for a general $\gamma(x^\perp)$. In the OPE in the exponent, we can now have derivatives of $\gamma$ and other geometric quantities, contracted with tensor operators,
\be
\partial_a \gamma\, \mathcal O^a+\partial_a \partial_b \gamma \,\mathcal O^{ab}+\partial_a \gamma\,\partial_b \gamma \,V^{ab}+\ldots
\ee
where $a=2, \ldots, d-1$ are the indices in the $x^\perp$ directions. Now, the key point is that the derivatives $\partial_a$ have twist one ($\gamma$ has zero twist), and to compensate, the operators must have twist less than $d-2$. This in turn strongly limits the possible operators that can enter the OPE. For instance, the vector operator $\mathcal O^a$, or any other symmetric tensor, has $\tau >d-2$. Hence these cannot appear. We are then left with antisymmetric tensors, which could have $\tau<d-2$. Consider the dominant contribution, coming from a rank-2 tensor contracted as $\partial_{[a}\gamma\,\partial_{b]}\gamma\,V^{[ab]}$. The operator needs to have twist exactly $d-4$. For a generic interacting CFT, this is not the case. Furthermore, since by unitarity $\tau >(d-2)/2$ in an interacting theory, we see that this can only occur for dimensions $d \ge 7$. Therefore, for $d \le 6$ these corrections are also absent. Intuitively, in the limit of small $\Delta x^-$ the two twist operators are very close to each other --their OPE is only sensitive to very UV information and is insensitive to the slow variation of the shape.

Having ruled out corrections from the nontrivial shape of $\gamma$, we now take the limit $\Delta x^+ \to \infty$, where the future boundary of a region $R_\gamma$ goes to infinity; see Fig. \ref{region}. In this case $\Delta x^+\,g(x^+/\Delta x^+) \to x^+$, which is the Rindler result ray by ray. Summing over $x^\perp$ obtains
\be
H_\gamma=2\pi \int d^{d-2} x^\perp\, \int_{\gamma(x^\perp)}^\infty d\lambda\, (\lambda-\gamma(x^\perp)) T_{\lambda \lambda}(\lambda,x^\perp)\,, \label{eq:DeltaH2}
\ee
up to an additive constant.

There is, however, an apparent paradox in taking the limit $\Delta x^+ \to \infty$. Indeed, for any fixed $\Delta x^+$  in the limit $\Delta x^-\rightarrow 0$ we have, for interacting theories in $d>2$, that $\Delta S = \Delta H$ and hence the relative entropy $S(\rho^1||\rho^0)=0$ \cite{Bousso:2014uxa}. This is because the products of operators in different copies of the replica space in the ellipsis in eq. (\ref{twist}) must have twist less or equal to $d-2$ to survive the $\Delta x^-\rightarrow 0$ limit. As we have seen this is not possible, unless $d=2$, in which case the stress tensor has zero twist and we could have products of the stress tensor in different copies, or in the case of free fields where products of two free fields of twist $(d-2)/2$ in different copies can have twist $d-2$, e.g. a product of two $\partial_+ \phi$ free fields.

However, when we take the limit $\Delta x^+\rightarrow \infty$ and perform the conformal map to the sphere, the null surface approaches the whole cone. The relative entropy must be generically non-zero for this case so that $\Delta S \neq \Delta H$ --either $\Delta H$ or $\Delta S$ should have a discontinuous limit. 

Let us first examine the limiting behavior for $\Delta H$. The question is whether the OPE (\ref{eq:OPE1}) restricted to operators in the original CFT (these are the ones that contribute to $\Delta H$) is well defined for large $x^+$. The expectation value of the operator at large $x^+$ is determined by the CFT on the cylinder as follows. From the mapping (\ref{eq:Mink-cyl}), $x^+ \to \infty$ corresponds to a point in the cylinder $t_c+\psi \to \pi/2$. We require $\langle \mathcal O_{cyl}\rangle$ to be finite there. The operator in Minkowski space is obtained by a Weyl rescaling, $\mathcal  O_{Mink} = \omega^\Delta\, \mathcal O_{cyl}$. From (\ref{eq:Weyl}), $\omega \sim 1/x^+$ as $x^+ \to \infty$, so we learn that
\be
\langle \mathcal O_{Mink}\rangle \sim \frac{1}{(x^+)^\Delta}\,.
\ee
Finally, plugging this behavior into (\ref{eq:OPE1}), we find a dependence on $x^+$ of the form $(x^+)^{s_k-\Delta_k -1}$. Since $\Delta_k - s_k \ge 0$, the behavior for large $x^+$ is bounded. Hence $\Delta \langle H\rangle$ has a smooth limit in taking the limits $x^+\rightarrow \infty$ and $x^-\rightarrow 0$, in any order, and (\ref{eq:DeltaH2}) is justified.

It must then be that the entropy jumps discontinuously for interacting theories in $d>2$. The reason to expect this is that the algebra of operators localized on the region of interest should also change abruptly. As it will be apparent below, the issue here is that of an order of limits. Indeed, for an interacting CFT, there remain no operators localized on a bounded surface in the null plane,\footnote{We mean here operators on the Hilbert space rather than field operators, which are operator valued distributions. These later of course can live on the null surface.} such as the strip we just considered. But taking the limit $\Delta x^+ \to \infty$, $\Delta x^-\rightarrow 0$ in some particular way to be specified below, the full QFT algebra should be regained. This can be seen more clearly on the cone. Let us denote by $\mathcal P_\gamma$ all the points on the null plane $\mathcal P$ to the future of $\gamma(x^\perp)$. This is conformally mapped to a surface $\mathcal C(\mathcal P_\gamma)$ that wraps the whole past light cone of a point with some boundary determined by $\gamma$. By causality, the algebra of operators should be the same as that on the region $R_\gamma$ that has this cone as future horizon. Hence we recover the algebra of all operators that can be localized on a volume of space-time. Since we have operators to distinguish the two states, we now expect $S(\rho^1||\rho^0) >0$.

To understand in more detail how this comes about, we study the behavior of smeared operators in a region of finite width $(\Delta x^+ , \Delta x^-)$. We smear the operator with a test function $\alpha(x)$, and compute the norm
\be\label{eq:norm}
||\mathcal O_\alpha |0\rangle||^2=\int d^dx\, d^dy\, \alpha(x)^* \alpha(y) \langle \mathcal O(x)^\dag \mathcal O(y) \rangle= \int \frac{d^d p}{(2\pi)^d}\,|\alpha(p)|^2 \langle \mathcal O(p)^\dag \mathcal O(p) \rangle\,.
\ee
For simplicity, let us focus on a scalar operator of dimension $\Delta$, in which case
\be
 \langle \mathcal O(p)^\dag \mathcal O(p) \rangle = \frac{1}{(p_+ p_-+ p_\perp^2)^{\frac{d}{2}-\Delta}}
\ee
in euclidean signature. 

First, if we want to localize an operator on a surface of constant time, $\alpha$ will be independent of $p_0$, and the possible UV divergence in the norm is of the form $||\mathcal O_\alpha|0\rangle||^2 \sim \int dp_0/p_0^{d-2\Delta}$. Hence in an interacting CFT we can have bounded operators on a spatial surface for $\Delta<\frac{d-1}{2}$. Next, consider the null limit, where the region is boosted to $\Delta x^- \to 0$. In this case, $\alpha$ is independent of $p_-$, and a potential UV divergence comes from $||\mathcal O_\alpha|0\rangle||^2 \sim \int dp_-/p_-^{d/2-\Delta}$. This would be finite for $\Delta <\frac{d-2}{2}$, which is below the unitarity bound and hence never satisfied.  This can also be extended to nonzero spin \cite{Bousso:2014uxa}. A complete calculation shows only some components of free fields such as $\partial_+\phi$  can be localized on the null surface. Thus there remain no operators in the null limit in the general case.

Now, let us consider the case of interest $\Delta x^- \to 0$ and $\Delta x^+ \to \infty$, keeping both finite to understand the limit. A finite $\Delta x^-$ is obtained from a test function with support on $|p_-|<1/\Delta x^-$; for the sake of the argument this can be chosen as a step function $\alpha(p) = \Theta((1/\Delta x^-)^2-p_-^2) \alpha(p_+,p_\perp)$. Then
\be
||\mathcal O_\alpha|0\rangle||^2 \sim \int^{1/\Delta x^-} \frac{dp_-}{p_-^{d/2-\Delta}}\,\int \frac{dp_+}{p_+^{d/2-\Delta}} \left(\int d^{d-2}p_\perp\,\alpha(p_+,p_\perp) \right)\,.
\ee
For test functions that smear the operator over a large $\Delta x^+$ we then find
\be
||\mathcal O_\alpha|0\rangle||^2 \sim \frac{1}{(\Delta x^- \Delta x^+)^{\Delta-(d-2)/2}}\,.
\ee
Thus, in the limit $\Delta x^- \to 0$ and $\Delta x^+ \to \infty$ with fixed $\Delta x^- \Delta x^+$, $\mathcal O_\alpha$  remains as a well-defined operator in the algebra. The presence of these operators explains the behavior of the entropy. We have $\Delta S = \Delta H$ if we take the limit $\Delta x^-=0$ first, but not if we take the limit $\Delta x^+=\infty$, $\Delta x^-=0$ with the product $\Delta x^- \Delta x^+$ fixed.

To end, we note that (\ref{eq:DeltaH2}) admits an alternative local form in terms of variations of surfaces. Calculating $\Delta H$ for the one-parameter family of curves
\be
\gamma(x^\perp, s)= \gamma(x^\perp,0)+s \dot \gamma(x^\perp)\,,
\ee
and taking two derivatives with respect to $s$, obtains
\be
\frac{d^2 }{ds^2}\Delta H_\gamma= 2\pi \int d^{d-2}x^\perp\,(\dot \gamma(x^\perp))^2\,T_{++}(x^-=0, x^+=\gamma(x^\perp), x^\perp)\,.
\ee
The integral over the null direction has canceled in taking the second variation with respect to $s$. For a small variation of the state, $\Delta S = \Delta H$, and we conclude that the QNEC \cite{Bousso:2015mna, Bousso:2015wca} is saturated for small deformations of the state in arbitrary CFTs. The saturation condition was obtained for free fields and holographically by \cite{Koeller:2017njr}.

\subsection{Modular Hamiltonians on the light cone}

Let us compute the general form of the modular Hamiltonian on the light cone. It is instructive to start with the expression on a finite strip (\ref{eq:strip}) rather than the one on semi-infinity regions (\ref{eq:DeltaH2}). 
The function $g(x)$ for $x\in (0,1)$ is dimensionless, symmetric under $x\rightarrow (1-x)$, and we expect it approaches $x$ for small $x$ and $(1-x)$ for $x\rightarrow 1$.  

We write (\ref{eq:strip}) as an integral over the transverse directions on the null surface of a quantity depending on the null interval 
\begin{equation}
dH=2\pi dA_\perp \,\int_{\lambda_0}^{\lambda_1} d\lambda\, \Delta \lambda\, g(\lambda/\Delta \lambda) \,T_{\mu\nu}(\lambda,y)\frac{dx^\mu}{d\lambda} \frac{dx^\nu}{d\lambda}\,, \label{qq3}
\end{equation}
where $\lambda$ is any affine parameter along the null interval, and $\Delta \lambda=\lambda_1-\lambda_0$.  
Let us analyse how this quantity transforms under conformal transformations.  
We first make a coordinate transformation such that the Minkowski metric in some new coordinates $x^\mu$ reads
\begin{equation}
\tilde{g}_{\mu\nu}(x)=\omega^2(x) \eta_{\mu\nu}\,.
\end{equation}
This step is just a coordinate transformation which does not change the expression of $dH$ in the null interval, provided we use the metric $\tilde{g}_{\mu\nu}$ and the corresponding $\tilde{T}_{\mu\nu}$ in the calculation in these coordinates
\begin{equation}
dH=2\pi d\tilde{A}_\perp \,\int_{\tilde{\lambda}_0}^{\tilde{\lambda}_1} d\tilde{\lambda}\, \Delta \tilde{\lambda}\, g(\tilde{\lambda}/\Delta \tilde{\lambda}) \,\tilde{T}_{\mu\nu}(\tilde{\lambda},y)\frac{dx^\mu}{d\tilde{\lambda}} \frac{dx^\nu}{d\tilde{\lambda}}\,. \label{qq1}
\end{equation}

Then we can map this problem by a conformal transformation eliminating the pre-factor $\omega^2$ in the metric (and hence the metric will turn to be $\eta_{\mu\nu}$) and at the same time changing the stress tensor which transforms as  
\begin{equation}
T_{\mu\nu}= \omega^{d-2} \tilde{T}_{\mu\nu}\,.   
\end{equation}
Due to the change in the metric the element of transversal area changes as
\begin{equation}
dA_\perp=\omega^{-(d-2)} d\tilde{A}_\perp\,.
\end{equation}
The previous affine parameter is not an affine parameter of the new (Minkowski) metric, and we have a change \cite{Wald:1984rg} 
\begin{equation}
d\lambda = c\, \omega^{-2} d\tilde{\lambda}
\end{equation}
with an arbitrary constant $c$. 

Using all this in eq. (\ref{qq1}) we get 
\begin{equation}
dH=2\pi dA_\perp \,\int_{\lambda_0}^{\lambda_1} d\lambda\, \Delta \lambda\,\,\left[\frac{\int_{\lambda_0}^{\lambda_1} d\lambda^\prime \omega^2(\lambda^\prime)}{(\lambda_1-\lambda_0)\omega^2(\lambda)}g\left(\frac{\int_{\lambda_0}^{\lambda} d\lambda^\prime \omega^2(\lambda^\prime)}{\int_{\lambda_0}^{\lambda_1} d\lambda^\prime \omega^2(\lambda^\prime)}\right) \,\right]\, T_{\mu\nu}(\lambda,y)\frac{dx^\mu}{d\lambda} \frac{dx^\nu}{d\lambda}\,.
\end{equation}
The expression in brackets is the new function $g$ which in general will be changed by the conformal transformation. 
A general special conformal transformation has 
\begin{equation}
\omega(x)=(1+2 (x\cdot c)+(x\cdot x)(c\cdot c))^{-1}\label{es}\,,
\end{equation}
with $c$ an arbitrary constant vector. 
We need the expression of $\omega(x)$ along a null line which can be parametrized as $x=x_0+\eta \lambda$, with $\eta$ a fixed null vector. Plugging this into (\ref{es}) we get
\begin{equation}
\omega(\lambda)=(c_0+c_1 \lambda)^{-1} \label{es2}\,,
\end{equation}
with $c_0$ and $c_1$ two constants. Then we can change affine parametrization such that the origin is at a point of the present coordinate system $x^\mu$. This point will be  the tip of the null cone.  With this parametrization we have  
\begin{equation}
\omega(\lambda)=\lambda^{-1} \label{es3}\,.
\end{equation}
Using this we obtain
\begin{equation}
dH=2\pi dA_\perp \,\int_{\lambda_0}^{\lambda_1} d\lambda\, \Delta \lambda\,\,g^*((\lambda-\lambda_0)/\Delta \lambda)\, T_{\mu\nu}(\lambda,y)\frac{dx^\mu}{d\lambda} \frac{dx^\nu}{d\lambda}\,, 
\end{equation}
with the new function $g^*$ given by
\begin{equation}
g^*(u)=\frac{((1-r)u +r )^2}{r} g\left(\frac{u}{(1-r)u +r}\right)\,,
\end{equation}
where $r=\lambda_0/\lambda_1$ is a number between $0$ and $1$.  Note $g^*$ is not symmetric under reflection for $r\neq 0,1$. If $\lambda_1\sim\lambda_0\gg \lambda_1-\lambda_0$, which corresponds to $r\rightarrow 1$, we get $g^*(u)\rightarrow g(u)$ as
 expected, since for a null interval far away from the tip of the cone the surface is approximately planar and $g$ is not modified. For the opposite case, $\lambda_0\rightarrow 0$, $r\rightarrow 0$, we get $g^*(u)\rightarrow u(1-u)$ for any $g$, using  $g(u)\sim u $ for small $u$.
 Hence for a region on the null cone containing the tip we have, using polar coordinates $\lambda,\Omega$ on the cone,
\begin{equation}
 H_\gamma=2 \pi \int d\Omega \,\int_0^{\gamma(\Omega)} d\lambda \,   \lambda^{d-1} \,\frac{\gamma(\Omega)-\lambda}{\gamma(\Omega)} \,T_{\lambda\lambda}\,, 
\label{dd}
\end{equation} 
where $T_{\lambda\lambda}=T_{\mu\nu}(\lambda,y)\frac{dx^\mu}{d\lambda} \frac{dx^\nu}{d\lambda}$, $\lambda$ is an affine coordinate over the null rays, $\lambda=0$ is the tip of the cone, and  $\gamma(\Omega)$ is a function of the angular coordinates giving the length of the null intervals.\footnote{We thank M. van Raamsdonk for pointing out a typo in the modular Hamiltonian (\ref{dd}) in a previous version of the paper.}
\subsection{Extension to massive deformations of CFTs}\label{subsec:massive}

So far we have considered the modular Hamiltonian for CFTs; we have established the result (\ref{eq:DeltaH2}) for an arbitrary shape in the null plane, and then used conformal transformations to find (\ref{dd}) for regions on the null cone. We now want to consider relevant deformations of a CFT, characterized by some mass parameter $m$.

The result for the cone (\ref{dd}) will not apply to massive theories, and we expect different behaviors depending on whether the size of the region is smaller or bigger than $1/m$. Intuitively, however, the formula for the null plane (\ref{eq:DeltaH2}) should apply to this case as well. Indeed, the analysis of Sec. \ref{subsec:shapes} showed how the OPE of twists operators is insensitive to IR deformations. In that case we focused on deformations of the shape, but we expect this to hold for more general relevant scalar perturbations of the CFT fixed point.

Consider then a perturbation
\be
S= S_{CFT}+\int d^d x \,g \,\mathcal O
\ee
by some operator of UV dimension $\Delta<d$. This triggers an RG flow at a scale $m \sim g^{1/(d-\Delta)}$. As $g \to 0$ we should recover (\ref{eq:DeltaH2}), and for a planar edge, $\gamma= \text{const}$, we have that the Rindler result is valid for massive theories. Hence in order to extend the OPE (\ref{eq:OPE1}) to the present case, we can only allow for positive powers of $g$ and/or derivatives of $\gamma(x^\perp)$. Possible shape corrections are absent as before, so let us focus on the effect of the deformation. In conformal perturbation theory, it can introduce new operators in the twist OPE of the form
\be
g^2\, \int d^{d-2} x^\perp \int dx^+\,\frac{1}{|\Delta x^- x^+|^{a_k}}\,(x^+)^{s_k-1} \mathcal O_{\Delta_k, s_k}+\ldots
\ee
Requiring this to be dimensionless fixes
\be
\Delta_k-s_k= d-2 -2 a_k -2(d-\Delta)\,.
\ee
Since $d-\Delta >0$ and $a_k \ge 0$ (because we take $\Delta x^- \to 0$), we have $\Delta_k-s_k <d-2$. But this is ruled out as discussed before. We thus conclude that the modular Hamiltonian (\ref{eq:DeltaH2}) for arbitrary shapes in the null plane is valid for massive deformations of CFTs.

\section{Some general mathematical properties of Modular Flows}
\label{sec:mate}

In the following sections we will need to use two powerful theorems about modular flows \cite{Borchers.unicidad}. The first one is an interesting property of full modular Hamiltonians and modular flows for regions which are moved into themselves by the modular flow 
of another region \cite{borchers1992cpt,wiesbrock1993half}. 

\bigskip

\noindent {\bf Theorem 1} (Half-sided modular inclusions): 
Suppose we have two von Neumann algebras ${\cal N}\subset {\cal M}$ with common cyclic and separating vector $|0\rangle$.  Let $U_{\cal M}(s)=e^{-i \hat{H}_{\cal M} s}$ be the unitaries implementing the modular flow of ${\cal M}$. Consider the case where the modular flow of ${\cal M}$ maps ${\cal N}$ in itself for all $s> 0$, 
\begin{equation}
{\cal N}(s)=U_{\cal M}(-s) {\cal N} U_{\cal M}(s)\subset {\cal N}\,, \hspace{2cm} s> 0 \,.\label{condi}
\end{equation}
If this property holds it is said one has a half-sided modular inclusion of algebras.\footnote{If the conditions (\ref{condi}) are assumed to hold for all $s\in R$, the problem has only trivial solutions.} In this case we have 

\bigskip

\noindent a) The family of algebras ${\cal N}_s$ with $s\in R$ is nested, ${\cal N}_{s_1}\subset {\cal N}_{s_2}$ for $s_1 >s_2$, with ${\cal N}_{-\infty}={\cal M}$ and ${\cal N}_0={\cal N}$. The modular flows of any member of this family of algebras move the other algebras of the family into themselves, in particular for the modular flow of ${\cal N}$
\be
U_{\cal N}(-u) {\cal N}_s U_{\cal N}(u)={\cal N}_{\frac{1}{2\pi}\log(1+e^{2 \pi u}(e^{2 \pi s}-1))}\,, 
\ee  
valid for all $s,u$ such that the argument of the logarithm in the right hand side is positive. 

\bigskip

\noindent b) The difference of the modular Hamiltonians for two included regions is a positive operator\footnote{This is always the case for included regions, whether half-sided or not. For an elementary derivation see \cite{Blanco:2013lea}.} $G=\hat{H}_{\cal M}-\hat{H}_{\cal N}\ge 0$, and we have the algebra of a two dimensional Lie group 
\begin{equation}
[\hat{H}_{\cal M},\hat{H}_{\cal N}]=i \,2 \pi (\hat{H}_{\cal M}-\hat{H}_{\cal N})=i 2 \pi G\label{commut}\,.
\end{equation}

\bigskip

\noindent c) The unitaries $T_{{\cal M},{\cal N}}(\tau)=e^{-i G \tau}$ generated by the positive operator $G$ are called modular translations.  We have that they map the algebra ${\cal N}$ into its modular translates 
\begin{equation}
T(-\tau) {\cal N}_s T(\tau)={\cal N}_{\frac{1}{2\pi} \log\left(e^{2 \pi s}+2 \pi \tau \right)} \label{ecu1}\,.
\end{equation} 
In particular, calling  ${\cal M}_\tau=T(-\tau) {\cal M} T(\tau)$, $\tau \ge 0$ to the translates of ${\cal M}$ we have
\be
{\cal M}_\tau={\cal N}_{\frac{1}{2\pi}\log(2 \pi \tau)}\,. \label{fijo}
\ee
Hence, the translations moves algebras into smaller ones for $\tau>0$, and ${\cal M}_0={\cal M}$, ${\cal M}_{(2 \pi)^{-1}}={\cal N}$. 

\bigskip

\noindent d) From the group we get the following relation between flows and translations
\be
U_{\cal M}(s)U_{\cal N}(-s)=T\left(\frac{1}{2\pi}(e^{2 \pi s}-1)\right)\,. \label{doce}
\ee

In order to better understand this theorem let us look at a simple example in QFT. First, in QFT any algebra of a region has the vacuum as a cyclic and separating vector. Cyclic means that acting on the vacuum with operators in the algebra we can approach any vector in the Hilbert space. Separating means that we cannot annihilate the vacuum with an operator in the algebra. Both of these conditions follow from the Reeh-Slieder theorem, see \cite{Haag:1992hx}. Let us then take the example of two wedges included into one another. In the null plane notation we can take $\gamma_1=0$ and $\gamma_2=c$, with any $c>0$, such that $\gamma_2\subset \gamma_1$. Considering that the modular flows of Rindler wedges are boosts, and that these act as dilatations on the null lines, it follows that this is a half-sided modular inclusion, and we get for the family of algebras
\be
\gamma_2(s)= c \, e^{2 \pi s}\,.   
\ee
These are just parallel wedges.
The modular translation generator is the difference between two boost generators $2\pi K_1$ and $2\pi K_2$ keeping fix $\gamma_1$ and $\gamma_2$ respectively. This difference is just a translation in the direction of the null ray 
\be
G= 2 \pi c \, P_\mu \xi^\mu\,,
\ee
which is a positive operator. All the above relations just follow from the Poincar\'e algebra. For example, translations on $\gamma_1$ are
\be
\gamma_1(\tau)=2 \pi c \, \tau\,,
\ee
in agreement with (\ref{fijo}). In fact, remembering the action of boosts and translations on parallel wedges is the simplest way to remember the half sided modular inclusion formulas.

\bigskip

The second theorem we need is one about the uniqueness of unitary flows with positive generator \cite{Borchers.unicidad,borchers1993symmetry,borchers1993modular}. 

\bigskip

\noindent {\bf Theorem 2}: Suppose we have a nested family of von Neumann algebras ${\cal N}_a$, $a\in R$, with ${\cal N}_a\subset {\cal N}_b $ for $a>b$, acting on a Hilbert space ${\cal H}$, with common cyclic and separating vector $|0\rangle$, and a one-parameter unitary group $T(a)$ with positive generator, leaving $|0\rangle$ invariant and translating the algebras
\be
{\cal N}_a=T(-a){\cal N}_0 T(a)\,. \label{trans}
\ee
Then any other one parameter unitary group with positive generator translating the algebras as in (\ref{trans}) and leaving $|0\rangle$ invariant coincides with $T(a)$. 

\section{Modular Hamiltonians on the null plane II. The algebra of $H_\gamma$}
\label{sec:algebra}

In this section we present a derivation of (\ref{hmodi1}) based on properties of half-sided modular inclusions and the uniqueness theorem of the previous section. In the process we uncover an infinite dimensional Virasoro algebra for $d$-dimensional field theories, acting on the null plane. We also comment on the extension to massive theories.

\subsection{Algebra of modular Hamiltonians on the null plane}\label{subsec:Halgebra}

As a warm-up, consider the set of wedges with edge contained in the null plane ${\cal P}$. If the edge of the wedge  passes though the origin, it can be labeled by the future-pointing null vector orthogonal to the $d-2$ dimensional edge $\gamma$. Call this null vector $\eta$, with the normalization $\eta\cdot \xi=1$. The corresponding modular operator is given by the boost generator leaving the wedge fixed,  
\be
\hat{H}_{(\eta)}=2 \pi J^{\mu\nu}\xi_\mu\eta_\nu \,,
\ee
where $J^{\mu\nu}$ is the Lorentz generator. This coincides with (\ref{hmodi1}) once $J^{\mu\nu}$ is written in terms of the stress tensor. 

Now let us consider wedges with the future horizon on the same plane, but not necessarily passing through the origin. Without loss of generality we can just translate a wedge through the origin in the direction parallel to $\xi$ by an amount $l \xi$. We have
\be
\hat{H}_{(\eta,l)}= e^{i P\cdot \xi\, l} \hat{H}_{(\eta)} e^{-i P\cdot \xi\, l}=  \hat{H}_{(\eta)}-2 \pi (\xi\cdot P) l\,,
\ee 
where we have used the Poincar\'e algebra
\be
[P_\lambda,J_{\mu\nu}]=i\left(g_{\nu\lambda} P_\mu-g_{\mu\lambda} P_\nu\right)\,.
\ee
Using
\be
[J_{\mu\nu},J_{\rho\sigma}]=i\left(g_{\nu\rho}J_{\mu\sigma}-g_{\mu\rho}J_{\nu\sigma}-g_{\nu\sigma}J_{\mu\rho}+g_{\mu\sigma}J_{\nu\rho}\right)\,,
\ee
we get the algebra for the modular Hamiltonians of wedges in the null plane ${\cal P}$,
\be
 [\hat{H}_{(\eta,l)},\hat{H}_{(\eta^\prime,l^\prime)}]= 2\pi i (\hat{H}_{(\eta,l)}-\hat{H}_{(\eta^\prime,l^\prime)})\,.\label{al}
\ee
This coincides with (\ref{commut}) for included planes, but extended also to intersecting planes.  

Now we want to generalize this result to more general regions on the null plane. Define the following operators by integrating on a null line parallel to $\xi$
\bea
P_{x^\perp}&=&\int d\lambda\, T_{\lambda\lambda}(\lambda,x^\perp)\,,\\
K_{x^\perp}&=&\int d\lambda\, \lambda\, T_{\lambda\lambda}(\lambda,x^\perp)\,.
\eea
These operators are space-like for different $x^\perp$, and their commutator has support only at coincident points in the coordinate $x^\perp$. Also this commutator has to commute with all fields space-like separated from the null line in question, and hence is an operator with support on this line. Another feature is that $P_{x^\perp}$ is invariant under translations along the null line while 
$K_{x^\perp}$ changes under these translations by the addition of a term proportional to $P_{x^\perp}$. Hence, the commutator is invariant under these translations. The commutator has dimension $2d-3$ and spin (boost eigenvalue) $1$. Therefore we can write, as distributions on the plane,
\be
[K_{x^\perp},P_{x^{\perp\,\prime}}]= \delta(x^\perp-x^{\perp\,\prime}) O_0(x^\perp)+ \partial_i \delta(x^\perp-x^{\perp\,\prime}) O_1^i(x^\perp)+\hdots \label{fof}
\ee
$O_0(x^\perp)$ is an operator localized on the line, with dimension $\Delta=d-1$, translational invariant, and spin $s=1$. Hence it has twist $\tau=\Delta-s=d-2$.  We can write it expanding in local operators on the line, 
\be
O_0=\int d\lambda\, \Phi(\lambda)+\int d\lambda\, \int d\lambda^\prime\, \Phi_1(\lambda)\Phi_2(\lambda^\prime)f(\lambda-\lambda^\prime)+\hdots\label{hdota}
\ee
The coordinate $\lambda$ has dimension $-1$ and spin $-1$, hence it has twist $0$. Then we need operators of twist exactly $d-2$. The twist of $T_{\lambda\lambda}$ is $d-2$ and, as we have recalled in Section \ref{sec:ope}, in general it will be the only operator with twist exactly $d-2$.     Hence we conclude the only possibility is $\Phi=c\, T_{\lambda\lambda}$. There are no other terms in (\ref{hdota}).

If we include derivatives as in the second term in (\ref{fof}) the situation is worse because the dimension of the necessary operators and their twist gets reduced. We do not have operators of twist less than $d-2$ to use as $O_1^i$. Antisymmetric representations multiplying higher derivatives of the delta function are absent by the same reason as in Sec. \ref{subsec:shapes}.
 
Then, calibrating the commutator using the one of momentum and boosts $[K_1,P_\xi]=-i P_\xi$ for the translation generator $P_\xi=\xi\cdot P$, we get
\be
[K_{x^\perp},P_{x^{\perp\,\prime}}]= - i  P_{x^{\perp}}\delta(x^\perp-x^{\perp\,\prime})\,. \label{dia}
\ee
Using this in the operators (\ref{hmodi1}) defined above we get a generalization of the algebra (\ref{al}) valid for all surfaces $\gamma$ 
\be
[\hat{H}_{\gamma_1},\hat{H}_{\gamma_2}]=2 \pi i (\hat{H}_{\gamma_1}-\hat{H}_{\gamma_2})\,.\label{gama}
\ee
It is not difficult to see that the algebra (\ref{gama}) does not admit non trivial central charges.  

\subsection{A Virasoro algebra in $d$ dimensions}

It is interesting to note that the argument that the commutator of $P_{x^\perp}$ and $K_{x^\perp}$ is a line integral of $T_{\lambda\lambda}$ also applies to more general operators of the form
\be
O_f(x^\perp)=\int d\lambda\, f(\lambda) \,T_{\lambda\lambda}(\lambda,x^\perp)\,,
\ee
with arbitrary non linear $f(\lambda)$.
Then we should have an algebra of the form $[O_f(x^\perp),O_g(x^{\perp\,\prime})]=\delta(x^\perp-x^{\perp\,\prime}) O_{h}(x^\perp)$, for some relation $h(f,g)$ between $f$ and $g$. 

Taking functions that are powers $\lambda$, this relation is fixed to be the one of Virasoro algebra for CFT in $d=2$ by the Lie algebra structure and dimensional analysis. In more detail, define
\be
L^n_{x^\perp} \equiv i \int d\lambda\,\lambda^{n+1}\, T_{\lambda\lambda}(\lambda, x^\perp)\,.
\ee
Matching dimensions and twists, and recalling that the commutator has to be proportional to a delta function in the transverse directions, obtains
\be
[L^m_{x^\perp},L^n_{y^\perp}]=\delta^{d-2}(x^\perp-y^\perp)\,(m-n)\,f(m,n)\,L_{x^\perp}^{m+n}
\ee
where we used the antisymmetry of the commutators, and the symmetric function $f(m,n)$ is so far undetermined. 
The Jacobi identity plus the value of $[L^{-1},L^m]$ that can be calibrated by the commutator with the momentum operator requires $f$ to be a constant. In this way, we arrive to the infinite-dimensional Virasoro algebra, ray by ray,
\be\label{eq:Virasoro}
[L^m_{x^\perp},L^n_{y^\perp}]=\delta^{d-2}(x^\perp-y^\perp)\,(m-n)\,L_{x^\perp}^{m+n}\,.
\ee
A central charge term is in principle also possible, however, the central charge is UV divergent on dimensional grounds \cite{Wall:2011hj}.\footnote{We thank Aron Wall for pointing this to us.} These operators would then transform the vacuum into infinite energy states. Still, there might be some renormalized version of the Virasoro symmetry that seems worth exploring. We leave a more detailed investigation of this very interesting point for a future work.  

\subsection{Positivity}

Our next step relies on the positivity of the operators $P_{x^\perp}$. This was proved in \cite{Faulkner:2016mzt} computing perturbatively the modular Hamiltonian of deformed Rindler space and the property of ordering of modular Hamiltonians for included regions \cite{Blanco:2013lea}. This property is equivalent to the averaged null energy condition for QFT and has interesting consequences \cite{Hofman:2008ar}. In this Section we are using this result, but note any independent derivation of the form of the modular Hamiltonians (\ref{hmodi1}) as the one in Section \ref{sec:ope} immediately gives this result as a consequence of the ordering of modular Hamiltonians for included regions.

The positivity of the operators $P_{x^\perp}$ implies they annihilate the vacuum. To see this consider the momentum operator that is an integral of these operators over the transversal direction. As the momentum annihilates the vacuum we have that
\be
\int dx^\perp\, \langle 0|P_{x^\perp}|0\rangle=0\,.
\ee
Since the positivity of $P_{x^\perp}$  implies the integrand is positive or zero, it must be that it is identically zero. For a positive operators $\langle 0|P_{x^\perp}|0\rangle=0$ implies 
\be
P_{x^\perp}|0\rangle=0\,.\label{rty}
\ee

From the ANEC we get the positivity of all operators  
\be
P_f=\int d^{d-2}x^\perp\, f(x^\perp)\int d\lambda\,  T_{\lambda\lambda}(\lambda,x^\perp)\,,\label{porf} 
\ee
where $f(x^\perp)>0$.  Eq. (\ref{rty}) also gives $P_f|0\rangle=0$.

\subsection{Action of $P_f$ on the algebras}

The operators $P_f$ act as generators of translations $\lambda\rightarrow \lambda+f(x_\perp)$ on the null surface. For a local operator $\phi(\lambda,x^\perp)$ on the null surface its action cannot be distinguished from the action of the momentum $f(x^\perp) P_\xi$ that has the same form in the null line where the operator is located. The part of $P_f$ away from this line will commute with the field. More formally, The commutator of $P_f$ with a field $\phi(\lambda,x^\perp)$ of dimension $\Delta$ and spin $s$ has to be localized and have dimension $\Delta+1$ and spin $s+1$, and generically we only have $\partial_{x^+}\phi(\lambda,x^\perp)$.\footnote{If the field is a descendant it can be a trade off between derivatives in the transversal directions and derivatives of $f$ in the commutator, i.e. $[P_f,\partial_{x_\perp^i}\phi]$ can contain $(\partial_{x_\perp^i} f(x^\perp)) \partial_{x^+}\phi$, as actually happens for parabolas, but still the evolution of the algebra is a translation in the $x^+$ direction.}  
On operators not located on the null surface the action of $P_f$ will be non local. 

The field operators on the null surface $\bar{\gamma}$ are moved by the translations $P_f$ to field operators in the null surface surface $\overline{\gamma+f}$. As operators in the bulk of the space-time region $R_\gamma$ commute with the field operators in the null surface $\bar{\gamma}$ it must be the case that the translated algebra $e^{i P_f} \gamma e^{-i P_f}$  commutes with the field operators in the null surface $\overline{\gamma+f}$ and this gives that this algebra is included in the one corresponding to the space-time region $R_{\gamma+f}$. The algebra $e^{iP_f}\bar{\gamma}e^{-i P_f}$ then includes the algebra of $R_{\overline{\gamma+f}}$. But it cannot be bigger since in that case some operators would not commute with the field operators in the null surface $\gamma+f$. In this way we arrive at
\be
e^{i P_f} \gamma e^{-i P_f}=\gamma+f\,. \label{eip}
\ee

Let us call $K_1$, $P_\xi$ as above to the boost and translation operators associated to the canonical wedge $W$ of $\lambda>0$. The modular Hamiltonian of this wedge is $H_W=2 \pi K_1$. Let us also call $U_\gamma(s)$ to the modular flows corresponding to $\gamma$  and $U_W(s)$ to the one of $W$. 

As the flow $U_W$ acts locally we can compute how it moves the regions $\gamma$,
\be
U_W(-s)\gamma U_W(s)=e^{2\pi s} \gamma\,. \label{miy}
\ee
Without loss of generality let us take $\gamma$ to lie in the region $\lambda>0$. Eq. (\ref{miy}) tell us $W$ and $\gamma$ are in the situation of half-sided modular inclusions described by Theorem 1. Then, there is a unitary one parameter group of modular translations $T_{W,\gamma}(\tau)$ with positive generator, that moves the regions proportional to $\gamma$ into themselves. According to (\ref{ecu1}) and (\ref{miy}) we get
\be 
T_{W,\gamma}(-\tau)\, \gamma \, T_{W,\gamma}(\tau)= (1+2 \pi \tau) \gamma\,.
\ee
From eq. (\ref{eip}), this is the same action as  
\be
e^{i \tau P_{2 \pi  \gamma}} \gamma e^{-i \tau P_{2 \pi \gamma}}= (1+2 \pi \tau )\gamma\,.
\ee
Then, these two one parameter groups of unitary operators, $T_{W,\gamma}(\tau)$ and $e^{-i \tau P_{2 \pi \gamma}}$,
 move the family of nested algebras $\kappa \gamma$, with $\kappa$ a constant into itself in the same way. 

Now we are in the position to use Theorem 2 because the generators $P_{2 \pi \gamma}$ are positive and annihilate the vacuum. Hence, we have the identity between the translations generated by the stress tensor and the modular translations
\be
G_{W,\gamma}=P_{2 \pi \gamma}\,.
\ee
From this and the fact that the modular translations are just the difference between the modular Hamiltonians, eq. (\ref{commut}), we finally get
\be
\hat{H}_\gamma=\hat{H}_W -P_{2 \pi \gamma}\,.
\ee 
This gives exactly the formula (\ref{hmodi1}) that we wanted to prove. The result (\ref{hmodi}) for $H_\gamma$ follows directly from this expression.
 With this identification all modular flows of surfaces $\gamma$ are local on the null surface, and  half-sided modular inclusions appear whenever $\gamma_1\ge \gamma_2$. The algebra of the generators (\ref{gama}) exactly matches the expected commutator for modular inclusions (\ref{commut}). 
 
It is interesting to comment on the physical meaning of Theorem 2 in the present context. We have two unitary flows that move some algebras of regions (as a whole) into algebras of other regions in the same way, and we wanted to show they have to be the same operators. In general this would not be correct: For example, we can add an internal symmetry generator such as the charge operator $Q$ to our generators written in terms of the stress tensor. As $Q$ commutes with the stress tensor and does not change the position of operators we get this new generator will also move the algebras in the prescribed way. However, the new generator will clearly not be positive if the original one was because $Q$ is unbounded below.    
 
Our discussion so far has been for CFTs, but the approach can be extended to massive deformations as in Sec. \ref{subsec:massive}. For this, we allow positive powers of the deformation parameter $g$ in (\ref{hdota}), but then recognize that this requires operators with twist below $d-2$, which cannot occur. Hence our result applies in conformal perturbation theory.

Finally, we note that spatial infinity of Minkowski space is an ordinary point in the cylinder, that can be considered the tip of a cone, and these infinite algebras of modular translations will keep this point at infinity fixed. It would be interesting to explore the relations between asymptotic symmetries at infinity that have been studied in the literature (see \cite{Strominger:2017zoo} for a recent review) with the infinite number modular symmetries for regions with future horizon lying on a common null cone that we are studying here.

\section{Modular flows on the null plane. Algebraic derivation}
\label{sec:algebraic}

In this section we give a general proof of the local action of modular flows on the algebras of the null surface,  based on algebraic methods. This approach is quite general, but we will here restrict to CFTs for some technical simplifications. The only input we use is the geometric modular flow of wedges and spheres in a CFT, apart from some standard assumptions about the algebra of intersections and unions of regions. It would be interesting to extend this to relevant deformations, as with the previous approaches.

If, for a range of modular parameter $s$, the modular flow of a region determined by $\gamma_1$ moves the algebra of another region $\gamma_2$  into algebras of regions in the following form, 
\be
U_{\gamma_1}(-s) \gamma_2 U_{\gamma_1}(s) =e^{2\pi s}(\gamma_2-\gamma_1)+\gamma_1\,,\label{standard0}
\ee   
we will say for brevity that the modular flow of $\gamma_1$ moves $\gamma_2$ ``in the standard way", for this range of $s$. 
Note that when the flow acts in a standard way we can compute its geometric action on each null line separately.  
In particular, we know that if $\gamma_1$ is a plane or a parabola, it will move any other region $\gamma_2$ in the standard way for all $s\in R$, according to the local modular flow of these surfaces in a CFT. 

\begin{figure}[t]
\begin{center}  
\includegraphics[scale=0.65]{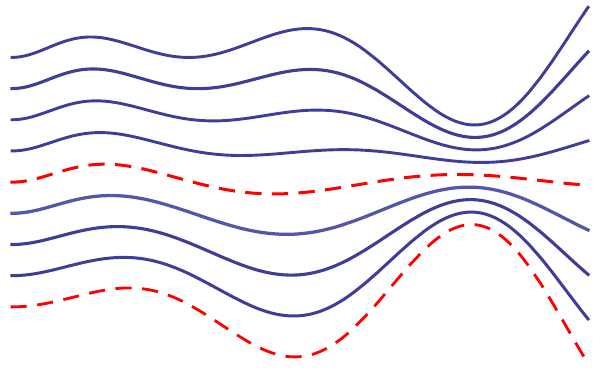}
\captionsetup{width=0.85\textwidth}
\caption{Family of curves that move among themselves by the modular translations corresponding to the two dashed curves. Here we have plotted (\ref{standardtrans}) for $\tau=l/4$, $l=0,1,\cdots 4$. The modular flows of these curves will also move the curves among themselves though with different parametrizations.}
\label{plano}
\end{center}  
\end{figure}

Then, suppose  $\gamma_1$  moves in the standard way for all $s\ge 0$ another region $\gamma_2$ above $\gamma_1$, that is, $\gamma_2>\gamma_1$. This will be a half sided modular inclusion for $s>0$. Then, there is a full family of regions contained above $\gamma_1$ that are the transform of $\gamma_2$ by the modular flow of $\gamma_1$, whose modular flow will move any other member of the family in the standard way.  We can label the family with the parameter $\tau$ of the modular translations corresponding to the pair $(\gamma_1, \gamma_2)$.  These, according to the algebra of half sided inclusions (\ref{commut}), can be written in terms of the modular flows of the two regions (\ref{doce}).
Then, using (\ref{standard0}), we can label the surfaces by the modular translation parameter
\be
\gamma_\tau=T_{\gamma_1,\gamma_2}(-\tau) \gamma_1 T_{\gamma_1,\gamma_2}(\tau) \,;
\ee
they correspond to the surfaces (see Fig. \ref{plano})
\be
\gamma_\tau=\gamma_1+ 2 \pi \tau (\gamma_2-\gamma_1)\,.
\ee
 We have $\gamma_{\tau=0}=\gamma_1$  and $\gamma_{\tau=(2\pi)^{-1}}=\gamma_2$, and the family includes all $\gamma_\tau$ for $\tau\in(0,\infty)$.   
There will be a half sided modular inclusion for any $\gamma_{\tau_1}<\gamma_{\tau_2}$ with $\tau_1\le \tau_2$ in the family, and  the modular translations of any of these ordered pairs will move the elements of the family in the ``standard way" for translations (see Fig. \ref{plano}), which we define to be:
\be
T_{\gamma_{\tau_1},\gamma_{\tau_2}}(-\tau)\gamma T_{\gamma_{\tau_1},\gamma_{\tau_2}}(\tau)=\gamma+ 2\pi\tau (\gamma_{\tau_2}-\gamma_{\tau_1})\,.\label{standardtrans}
\ee
Note that the modular flow parameter $s$ and modular translation parameter $\tau$ in transformations that move the algebras in the standard way, are constrained by the requirement that none of the surfaces of the family go below the original $\gamma_1$. The surfaces of the family extend from $\gamma_1$ to infinity, passing through $\gamma_2$.

If $\gamma_1$ moves $\gamma_2$ and $\gamma_3$ both in the standard way for a range of $s$, it will move the intersection 
\be\label{eq:int}
\gamma_2 \cap \gamma_3=\theta(\gamma_2-\gamma_3)\gamma_3+\theta(\gamma_3-\gamma_2)\gamma_2\,, 
\ee 
and the union
\be\label{eq:union}
\gamma_2 \cup \gamma_3=\theta(\gamma_2-\gamma_3)\gamma_2+\theta(\gamma_3-\gamma_2)\gamma_3\,,
\ee
 in the standard way for the same range of the modular parameter. This follows directly from the fact that modular flows are unitary transformations that preserve commutation relations, and that we are assuming that for algebras corresponding to geometrical regions the complementary region corresponds to the commutant algebra (Haag duality), the intersection of the algebras corresponds to the algebra of the intersection of regions, and the generated algebra of two regions on the null plane corresponds to the union of regions. 
 
 In the same way, if $(\gamma_1,\gamma_2)$ is an ordered pair in the half-sided situation whose modular flows (and hence the corresponding modular translations) move in the standard way two other regions $\gamma_3$, $\gamma_4$,  the intersection and union of these last two regions will be moved in the standard way by the modular translations of the pair.

Then let us take a surface $\gamma$ that admits a plane below it. We know that the modular flow of $\gamma$ moves in the standard way the modular translates of any plane below $\gamma$ to the corresponding surfaces above $\gamma$.  Taking an arbitrary number of different planes below $\gamma$ we can translate them above $\gamma$ and then make arbitrary intersections and unions of these surfaces, all of which will be moved in a standard way by the modular flow of $\gamma$. This gives a large family of regions above $\gamma$  that are moved in standard way by $\gamma$. However, there are some convexity constraints on the type of regions one can make by this procedure. It would be interesting to explore further this construction for Lorentz invariant theories. In order to have a more general result, we will use conformal invariance on top of Lorentz invariance. 
 The new powerful ingredient of the CFT is that we know now the modular flow corresponding to any parabolic surface (\ref{parabola}) will also move arbitrary regions in the standard way.

Then, we can take $\gamma$ and any parabolic region below $\gamma$. We can use the modular translations to pass the parabola above $\gamma$ to a family of surfaces that will be moved in the standard way by the modular flow of $\gamma$. These parabolas can be as tightly aligned with a null line as we want.  As we move any of the individual parabolas above $\gamma$ they soon become tightly aligned with the null ray pointing to the future of $\gamma$. We can take any number of these transformed parabolas as tightly aligned to null rays as we want, and do unions of them to approach any region above $\gamma$. 
This region will be moved in the standard way by the flow of $\gamma$. As a result, any $\gamma$ will move in the standard way any other region above it. The same can be said for regions below $\gamma$. The algebra of the modular Hamiltonians for arbitrary regions included in one another follows from the half-sided theorem and coincides with  
 (\ref{gama}) for this case. 

Finally, we prove an identity for modular Hamiltonians that in a certain sense tells they have to be understood as a sum over null lines as in (\ref{hmodi1}). This identity will be important for the next section when we discuss the Markov property.

 We take two intersecting regions $\gamma_1$ and $\gamma_2$. $\gamma_1$ will move $\gamma_1\cap \gamma_2$, that is above $\gamma_1$, in a standard way, and $\gamma_1\cup \gamma_2$ will move $\gamma_2$ in a standard way, towards the future of the plane. Both of these flows are half sided and their modular translation generators are positive. Further, the corresponding modular translations of these two pairs, $(\gamma_1,\gamma_1\cap \gamma_2)$ and $(\gamma_1\cup \gamma_2,\gamma_2)$, move any algebra $\gamma_3$ above $\gamma_1\cap \gamma_2$ in a standard way, eq. (\ref{standardtrans}), as follows from the fact that modular translations are products of modular flows, eq.(\ref{doce}).  
 
In fact, both modular transformations move $\gamma_3$ in identical manner: 
\bea
 T_{\gamma_{1},\gamma_1\cap \gamma_{2}}(-\tau)\gamma_3 T_{\gamma_{1},\gamma_1\cap\gamma_{2}}(\tau)=\gamma_3+ 2 \pi\tau (\gamma_1\cap \gamma_{2}-\gamma_{1})\nonumber \\
=\gamma_3+ 2 \pi \tau (\gamma_2-\gamma_1\cup \gamma_2)=T_{\gamma_{1}\cup \gamma_2,\gamma_2}(-\tau)\gamma_3 T_{\gamma_{1}\cup \gamma_2,\gamma_2}(\tau)\,.
\eea 
Both of these modular translations with positive generator push the algebra $\gamma_3$ into itself to the future. Then we can apply Theorem 2 of section \ref{sec:mate} to conclude that the translation unitary operators are identical, and their generators are in fact the same.
 Again, the physical interpretation here is that they move the algebras in a local way, and another different operator that would do the same job would differ by an internal symmetry transformation and would not be positive. 
 Therefore we have
\be
\hat{H}_{\gamma_1}-\hat{H}_{\gamma_1\cap\gamma_2}=\hat{H}_{\gamma_1\cup \gamma_2}-\hat{H}_{\gamma_2}\,. \label{markomenos}
\ee 
This is the general identity that we wanted to prove.

\section{The vacuum as a Markov state}
\label{sec:markov}

In this section we first review Markov states, and their basic properties, together with the derivation of the main identity. Then we generalize these knows results to be used with (\ref{markomenos}) and (\ref{markovsinhat}) derived in this paper. We establish the Markovian property of the field theory vacuum state, and go on to prove the strong super-additivity of the relative entropy.

\subsection{Markov states} 

Classically, three random variables $X$, $Y$, $Z$, form a Markov chain if the conditional probabilities satisfy $p(x|y,z)=p(x|y)$. This means that we do not learn more on $X$ by having a knowledge of $YZ$ than just by having knowledge of $Y$. This property coincides with saturation of strong subadditivity
\be
S(XY)+S(YZ)=S(XYZ)+S(Y)\,.
\ee 

A state that saturates strong subadditivity is called a Markov state, even in the quantum domain. In \cite{petz1988sufficiency,petz2007quantum} it was shown that this numerical equation is equivalent to a full operator equation:
\begin{equation}\label{inden}
S(A)+S(B)-S(A \cap B)-S(A \cup B)=0 
\Leftrightarrow 
 -\log \rho_A-\log \rho_B+\log \rho_{A\cap B}+\log \rho_{A\cup B}=0\,.
\end{equation}
Using the notation $A=(12)$, $B=(23)$, $A\cap B=(2)$, $A\cup B=(123)$, the Markov property is also equivalent to the following structure of the density matrix $\rho_{A\cup B} \equiv \rho_{123}$  \cite{hayden2004structure}. There exists a decomposition of the Hilbert space ${\cal H}_2$ as a direct sum of tensor products ${\cal H}_2=\oplus_k {\cal H}_L^k\otimes {\cal H}^k_R$ such that 
\be
\rho_{123}=\sum_k p_k\, \rho^k_{1\,L}\otimes \rho^k_{R\,3}\,,\label{chara}
\ee
where $p_k$ are probabilities. There are no other correlations between the subsystems $(1)$ and $(3)$ other than the ones mediated by $(2)$. 
Note that this expression gives $\rho_{13}=\sum_k p_k \rho^k_1\otimes \rho^k_3$. Then there is no entanglement between $(1)$ and $(3)$, while there could be classical correlations if the sum contains more than one different term. Conversely, any separable $\rho_{13}$ admits a Markov extension. 

Another striking quantum information theoretic characterization arising from (\ref{chara}) is that for a Markov state we can reconstruct the state in $A\cup B$ from the knowledge of the state in $A$ and $B$ \cite{hayden2004structure}. This property is analogous to the one of classical Markov chains.

Let us briefly review the derivation of (\ref{inden}) (and of strong subadditivity) as presented in   \cite{petz2007quantum}. This will allow us to introduce the necessary tools for our generalizations of this result. 

Taking expectation value of the right hand side of (\ref{inden}) on the global state we get the left hand side. To show the opposite implication, consider the matrix  
\be
\exp(\log \rho_A -\log \rho_{A\cap B}+\log \rho_{B})=\lambda \, \omega\,.
\ee
This is a positive operator that can be written as $\lambda \,\omega$ with $\omega$ a density matrix and $\lambda>0$ a number. A simple computation gives
\be
S(A)+S(B)-S(A\cap B)-S(A\cup B)=S(\rho_{A\cup B}||\,\omega)-\log \lambda.  \label{esp}
\ee 

Following \cite{petz2007quantum} we use the Golden-Thomson-Lieb inequality \cite{LIEB1973267} 
\be
\textrm{tr} \exp(X+Y+Z)\le \int_0^\infty dt\, \textrm{tr} \left((t+e^{-X})^{-1} e^Y (t+e^{-X})^{-1} e^Z\right)\,,  \label{gtl}
\ee
which holds for any three hermitian matrices $X$, $Y$, $Z$.
Taking $X=-\log \rho_{A\cap B}$, $Y=\log \rho_{A}$, $Z=\log \rho_B$, it follows that
\bea
\lambda &\le&  \int_0^\infty dt\, \textrm{tr} \left((t+\rho_{A\cap B})^{-1} \rho_{A} (t+\rho_{A\cap B})^{-1} \rho_{B}\right) \nonumber \\
&=& \int_0^\infty dt\, \textrm{tr} \left((t+\rho_{A\cap B})^{-1} \rho_{A\cap B} (t+\rho_{A\cap B})^{-1} \rho_{A\cap B}\right)=\textrm{tr}\rho_{A\cap B}=1\,.\label{dirac0}
\eea
In the first step we have taken the trace over the part of $A$ and $B$ that is not in the intersection. In the second step we computed the integral explicitly. 

 Hence, $\log(\lambda)\le 0$ and in (\ref{esp}) we get strong subadditivity for the entropies on the left hand side from positivity of the relative entropy on the right hand side. Now, for the left hand side in (\ref{esp}) to vanish, we need both $\lambda=1$ and $S(\rho_{A\cup B}||\,\omega)=0$, since the relative entropy and $-\log(\lambda)$ are positive. Then, the density matrices  $\rho_{A\cup B}$ and $\omega$ are the same, and 
\be
\rho_{A\cup B}=\exp(\log \rho_A -\log \rho_{A\cap B}+\log \rho_{B})\,,
\ee 
which is equivalent to the right hand side of (\ref{inden}).

\subsection{Generalizations}

We will need a slightly more general relation than (\ref{inden}), which is more adapted to the continuum limit for QFT applications. In QFT, the modular Hamiltonian $H$ is well defined as a generator of unitary transformations $U(\tau) = e^{-i H\, \tau}$ acting on the operators. These modular transformations determine $H$  up to an additive constant. This constant is cutoff dependent because of the cutoff dependence of the normalization of density matrices in the continuum limit. Then we have that, for a generic space-time region $X$, 
\be
H_X=-\log \rho_X+c_X\,,
\ee
with $c_X$ an undetermined constant. We have found above that certain combinations of modular Hamiltonians vanish, but the constants $c_X$ for these operators are non zero. Then we need to replace (\ref{inden}) with
\be
S(A)+S(B)-S(A\cap B)-S(A\cup B)=0 \Leftrightarrow H_A+H_B-H_{A\cap B}-H_{A\cup B}=0 \,.\label{inden1}
\ee 

A priori this is not the same as (\ref{inden}). The equivalence follows if we can show that  
\be
-\log \rho_A-\log \rho_B+\log \rho_{A\cap B}+\log \rho_{A\cup B}=k \implies k=0\,,\label{ade}
\ee
where $k=-c_A-c_B+c_{A\cap B}+c_{A\cup B}$ is a constant. 

First note that taking the expectation value on the left hand side of (\ref{ade}) in the state $\rho_{A\cup B}$ and using strong subadditivity we get
\be
k\ge 0\,.
\ee
Now, we also have from the left hand side in (\ref{ade}) that
\be
\textrm{tr}\exp(\log \rho_{A\cap B}+\log \rho_{A\cup B}-\log \rho_{A})=e^{k} \, \textrm{tr}\rho_{B}=e^{k}\,. 
\ee  
Taking $X=-\log \rho_{A}$, $Y=\log \rho_{A\cap B}$, $Z=\log \rho_{A\cup B}$, in (\ref{gtl}) we get
\bea
e^k&\le &  \int_0^\infty dt\, \textrm{tr} \left((t+\rho_A)^{-1} \rho_{A\cup B} (t+\rho_A)^{-1} \rho_{A\cap B}\right) \nonumber\\
&=& \int_0^\infty dt\, \textrm{tr} \left((t+\rho_A)^{-1} \rho_{A} (t+\rho_A)^{-1} \rho_{A\cap B}\right)=\textrm{tr}\rho_{A\cap B}=1\,.\label{dirac}
\eea
In the first step we have used that the state $\rho_{A\cup B}$ evaluates the expectation value of an operator on the algebra $A$, and for this, it is sufficient to replace it by $\rho_A$. In the second step we have done the integration in $t$ explicitly. This implies $k\le 0$ and then $k=0$ as we wanted to prove. Hence, (\ref{inden1}) follows from (\ref{inden}). 

The expression (\ref{hmodi}) we have obtained for the modular Hamiltonians $H_\gamma$ in terms of the stress tensor leads directly to the Markov property for any two regions on the null plane or cone in a CFT, or on the null plane for field theories with relevant perturbations.  In this expression the modular Hamiltonians are not normalized as $H_\gamma=-\log \rho_\gamma$. Rather, since the vacuum expectation value of the stress tensor is zero, they are normalized as $H_\gamma=-\log \rho_\gamma -S_\gamma$, with $\langle 0| H_\gamma| 0\rangle=0$. The Markov property then implies that these constants add up to zero as in (\ref{ade}). As these constants are here the entropies, this already implies the left hand side of (\ref{inden1}). The vacuum state in a CFT is Markovian for any regions on the null cone or the null plane. 
 
\bigskip

The result 
\be
H_A+H_B-H_{A\cap B}-H_{A\cup B}=0
\ee 
for modular Hamiltonians on the null plane follows from the explicit expression of the modular Hamiltonian in terms of the stress tensor. If we want to use our general algebraic result of Section \ref{sec:algebraic}, we must derive the Markovian property from the vanishing combination of full modular Hamiltonians (\ref{markomenos}). We need to show that 
 \be
\hat{H}_A+\hat{H}_B-\hat{H}_{A\cap B}-\hat{H}_{A\cup B}=0\Rightarrow S(A)+S(B)-S(A\cap B)-S(A\cup B)\,.\label{ulty}
\ee
However, here we have an obstacle in trying to use our finite dimensional methods to deal with this implication.
We want to set the left hand side equal to zero and at the same time have a pure global state in finite dimensions. While this is possible in QFT, in finite dimensional Hilbert spaces with global pure state, the number $n_X$ of zero eigenvalues of the density matrix on some subsystem $X$ and the one of its complement satisfies 
\be
n_X-n_{\bar{X}}=d_{X}-d_{\bar{X}}\,,
\ee 
where $d_X$ is the dimension of the Hilbert space ${\cal H}_X$. 
This is easily shown using the Schmidt decomposition. Then while we can make $n_X$ and $n_{\bar{X}}$ both zero when the dimensions are equal, for our regions $A$, $B$, $A\cap B$, $A\cup B$, and their complements, we cannot make the number of zeros of all density matrices vanish at the same time. If a density matrix has a zero then its logarithm (and its modular Hamiltonian) has an infinite eigenvalue, and we need to use all modular Hamiltonians to define the left hand side of (\ref{ulty}). Note however that the zeros of the density matrices pose no problem for the entropies on the right hand side of (\ref{ulty}). 

To deal with this technical problem and remain in finite dimensions we will impose a cutoff to the purity of the global state, that we remove at the end.      

In order to proceed, let us first for convenience change the notation for the regions. We call $A=(12)$, $B=(23)$, $A\cap B=(2)$, $A\cup B=(123)$, and $\bar{A}=(34)$, $\bar{B}=(14)$, $\overline{A\cap B}=(134)$, $\overline{A\cup B}=(4)$. If our pure global state is $\rho_{1234}=|0\rangle\langle 0|$ we take as global state instead
\be
\tilde{\rho}_{1234}=(1+\epsilon)^{-1}\left( |0\rangle\langle 0|+\epsilon \frac{1}{d_{1234}}\right)\,,
\ee
with small $\epsilon$. We impose the left hand side of (\ref{ulty}) to $\tilde{\rho}_{1234}$ instead of $\rho_{1234}$. A nice thing about this regularization is that we have
\be
\tilde{\rho}_X= (1+\epsilon)^{-1}\left(\rho_X+\epsilon \frac{1}{d_{X}}\right)\,,
\ee 
and, of course, for small enough $\epsilon$, the entropies go to the ones of the density matrix $\rho_X$,
\be
\tilde{S}(X)\sim S(X)-\epsilon \log(\epsilon) \frac{n_X}{d_X}+{\cal O}(\epsilon)\,.
\ee 
  
The left hand side of (\ref{ulty}) implies, on our regularized state,
 \be
-\log \tilde{\rho}_{12}-\log \tilde{\rho}_{23}+\log \tilde{\rho}_{2}+\log \tilde{\rho}_{123}=k_{13}=-\log \tilde{\rho}_{34}-\log \tilde{\rho}_{14}+\log \tilde{\rho}_{134}+\log \tilde{\rho}_{4} \,.\label{ultym1}
\ee
Now $k_{13}$ can be an operator instead of a constant as in (\ref{ade}). 
From the form of the two sides of this equation it is not difficult to see that $k_{13}$ commutes with all operators in $(2)$ and $(4)$. We also note that (\ref{ultym1}) is invariant under interchange of $(1)\leftrightarrow(3)$ and $(2)\leftrightarrow (4)$.

Taking expectation values on the global state we get from strong subadditivity
\be
\langle k_{13}\rangle\ge 0\,.
\ee
We also have from the same calculations that lead to (\ref{dirac}),
\be
\textrm{tr}\exp(\log \tilde{\rho}_{2}+\log \tilde{\rho}_{123}-\log \tilde{\rho}_{12})=\textrm{tr}\,e^{\log \tilde{\rho}_{23}+k_{13}}\le 1\,. 
\ee 
In order to proceed we need to use the representation \cite{petz2007quantum}
\be
\textrm{tr}\,e^{\log \rho+Q}=e^{\max_{\rho^\prime}(\textrm{tr}(Q\rho^\prime)-S\left(\rho^\prime||\rho)\right)}\,,\label{marte}
\ee
where the maximum is over all density matrices $\rho^\prime$. Taking $Q=k_{13}$, $\rho=(1/d_1) \otimes \tilde{\rho}_{23}$ and $\rho^\prime=\tilde{\rho}_{123}$ in (\ref{marte}) we have
\be
e^{\langle k_{13}\rangle} \,e^{\tilde{S}_{23}-\tilde{S}_{123}}=e^{\langle k_{13}\rangle} \,e^{S_{23}-S_{123}+{\cal O}(\epsilon\log \epsilon)}\le\textrm{tr}\,e^{\log \tilde{\rho}_{23}+k_{13}}\le 1\,.\label{m1}
\ee
Interchanging $2\leftrightarrow 4$, and taking into account that $S_{34}=S_{12}$ and $S_{143}=S_2$ we have
\be
e^{\langle k_{13}\rangle} \,e^{S_{12}-S_{2}+{\cal O}(\epsilon \log{\epsilon})}\le 1\,.\label{m2}
\ee
Multiplying (\ref{m2}) and (\ref{m2}) we have
\be
 e^{2\,\langle k_{13}\rangle} \,e^{S_{12}+S_{23}-S_{2}-S_{123}+{\cal O}(\epsilon \log \epsilon)}\le 1\,.\label{simpli}
\ee
From strong sub-additivity $\langle k_{13}\rangle\le {\cal O}(\epsilon \log(\epsilon))$, and hence 
\be
\langle k_{13}\rangle={\cal O}(\epsilon \log(\epsilon))\,.\label{mara1}
\ee
 Using (\ref{simpli}), this gives 
\be
S_{12}+S_{23}-S_{2}-S_{123}={\cal O}(\epsilon \log(\epsilon))\,,\label{mara}
\ee
and the state is Markovian as we remove the cut-off.

A startling point  arises when we look back at equations (\ref{m1}) and (\ref{m2}), equipped with the knowledge of (\ref{mara1}). Together with strong sub-additivity they give that all involved entropies $S_{12}$, $S_{23}$, $S_{123}$, $S_{2}$ are equal, up to errors of the order $\epsilon \log{\epsilon}$. Hence, at least in this finite dimensional case, and removing the $\epsilon$ cut-off faster than taking the large dimension limit, the left hand side of (\ref{ulty}) seems to be stronger than the Markov property. Apparently this way of going to the continuum limit does give the strong sub-additive saturation correctly but does not leave enough liberty to the entropies to encompass the values of the entropies of regions on the light cone in a CFT.

However, we want to speculate that another interpretation might actually be possible. Perhaps this is suggesting there should be an (unknown) regularization of the entropy in QFT such that all entropies of regions on the cone are equal in a CFT. This may not be such a bold statement as it seems at a first sight since, for example, most of what produces a different entropy for different regions is the area term, which is non universal, and changes with the cut-off. We also recall that all these regions are mapped unitarily to each other by some modular flows with geometric action on the null cone and which leave the vacuum invariant. If we impose a cutoff that changes according with these transformations the entropies of the transformed regions will be equal. In fact, following a calculation similar to the one in \cite{Casini:2012ei}, that will be presented elsewhere, one can show for example that the universal finite term for any region on the light cone in $d=3$ is actually always the same number $F$, that gives the universal part of the entropy of a circle.            

In any case, it is important to recognize that this uncertainty about how to take the continuum limit for the identity of full modular Hamiltonians does not affect our previous proof for the identity of the modular Hamiltonians. It is always possible to go to the continuum with the state in $\rho_{A\cup B}$ having no zeros for the density matrices involved in this relation. Hence, even if we know that both identities hold in the CFT, we can always use the one on the modular Hamiltonians restricted to subregions of $A\cup B$ to prove saturation of strong sub-additivity. 
 
\subsection{Strong super-additivity of relative entropy} 
 
For applications to QFT we want to write consequences from the Markov property in terms of quantities that have a well-defined continuum limit. We have already noted that the right hand side of (\ref{inden}) can be written in terms of the modular Hamiltonians with unspecified normalizations, or in terms of the full modular Hamiltonians. Next we want to replace entropies by differences of entropies between different states, or by relative entropies. These are finite in the continuum limit. 
If $\rho^0$ is Markovian, from SSA for the (arbitrary) state $\rho^1$,  we get
\begin{equation}
\Delta S(A)+\Delta S(B)-\Delta S(A\cap B)-\Delta S(A\cup B)\ge 0 \,.
\end{equation}
Because of (\ref{inden}), we have $H_A+H_B-H_{A\cap B}-H_{A\cup B}=0$ for $\rho^0$. Then we can use $\Delta H_A+ \Delta H_B-\Delta H_{A\cap B}-\Delta H_{A\cup B}=0$ and subtract the above equation to obtain
\begin{equation}\label{SSupA}
S(\rho_1^A||\rho_0^A)+S(\rho_1^B||\rho_0^B)-S(\rho_1^{A\cap B}||\rho_0^{A\cap B})-S(\rho_1^{A\cup B}||\rho_0^{A\cup B})\le 0
\end{equation}
where again $\rho^0$ is Markovian and $\rho^1$ is arbitrary.
This is the strong super-additivity of relative entropy. It does not hold for general states $\rho^0$.  Here we obtain strong super-additivity  as a consequence of Markov property of the state $\rho^0$.\footnote{Some other sufficient conditions for its applicability  were studied in \cite{petz2007quantum} in a more general context. They have the form of quite restrictive constraints on the state $\rho^0$. In the present case involving tensor product of Hilbert spaces, these conditions ask for the state $\rho^0$ to be of a product type $\rho^0 = \omega_1\otimes \omega_2 \otimes \omega_3$,  \cite{petz2007quantum}, what is not the case for the vacuum of a QFT.} 

In an upcoming work, we will show how this property leads to an entropic proof of the $a$-theorem in four dimensions, and unifies previous results on the monotonicity of the RG.

\section{Conclusion}   
We have shown that the modular Hamiltonians for spatial regions having  arbitrary boundary lying on a null plane in Minkowski space have a universal form given by an integral of the stress tensor. This is of the form of the Rindler result ray by ray, and there are no cross terms mixing the different null rays. We proved this both using an OPE expansion of twisting operators and computing the algebra of the modular Hamiltonians and using general results about modular flows. These modular Hamiltonians do not have a local form when expressed in Cauchy surfaces other than the null surface.   For CFT's an equivalent result holds for arbitrary regions having boundary on a light cone. 

The corresponding modular flows move operator algebras geometrically on the null plane (or cone) at the velocity given by the Rindler flows on each ray. The flow of field operators outside the null surface should be very non local in general, except for the particular cases corresponding to conformal symmetry generators. We could prove this result for a CFT in  full generality using the theory of half-sided modular inclusions of algebras. The modular Hamiltonians form a fixed infinite dimensional Lie algebra that is in accordance with this theory.   

This results leads to a Markov property of the vacuum state over this particular type of regions. This is  an identity for modular Hamiltonians involving two regions, their intersection and union. As a consequence we have the saturation of the strong sub-additive inequality for entropies. We used this to establish the strong super-additivity of relative entropy between a Markov state and an arbitrary state. This is an important element in a new proof of the a-theorem for renormalization group flows in $d=4$ presented in \cite{Casini:2017vbe}.

A surprising consequence of the Markov property in this case is that we can reconstruct the state on the union $\gamma_1\cup \gamma_2$ of two regions on the null plane from the knowledge of the state on each of the two. In the light cone picture these regions have to share a neighbourhood of the tip of the cone. This suggest that entanglement has a simpler structure than the one we could have naively expected. 

The specific form of modular Hamiltonians on the null plane involving the stress tensor generalizes the result for free fields in \cite{Wall:2011hj}, and shows the proof of the generalized second law in the weak gravity limit in that paper is valid for interacting models as well.     

As an important by-product of our calculation of modular Hamiltonians we found that the infinite dimensional ``symmetries"
given by modular flows on the cone could be enlarged to a bigger algebra of the Virasoro type on each null ray. These have divergent central charges, are not given by conserved charges produced by a local current, but we naturally wonder of their possible uses to understand  higher dimensional CFTs. In particular we would like to understand if the central charge can be renormalized in a useful way.

The infinite algebra of modular Hamiltonians has seemingly connections to the infinite groups of asymptotic symmetries of space-time that have been discussed in the literature. It would be interesting to establish a connection between these symmetries and modular flows.

\section*{Acknowledgments}
H.C. thanks Marina Huerta for discussions and encouragement, and Aron Wall for comments on an earlier version of the manuscript. E.T. thanks the ICTS/Prog-infoasi/2016/12, Bangalore, and the Simons Center in Stony Brook for hospitality during part of the realization of this work.   
This work was partially supported by CONICET, CNEA, and Universidad Nacional de Cuyo, Argentina. H.C. acknowledges an ``It From Qubit" grant of the Simons Foundation. 
G.T. is also supported by ANPCYT PICT grant 2015-1224.


\bibliography{EE}{}
\bibliographystyle{utphys}

\end{document}